\documentclass[%
twocolumn,
groupedaddress,
longbibliography,
amsmath,amssymb,
 aps,
prb,
]{revtex4-1}
\usepackage{xcolor}
\usepackage{enumerate}
\usepackage{physics}
\usepackage[all]{xy}
\usepackage[T1]{fontenc}
\usepackage{amsmath,amsfonts,amssymb, braket}
\usepackage{wasysym}
\usepackage{comment}
\usepackage{enumerate}
\usepackage{graphicx}
\usepackage{mathtools}
\usepackage{ifthen}
\usepackage{tensor}
\usepackage{braket}
\usepackage{standalone}
\usepackage{color}
\definecolor{myred}{RGB}{232,102,102}
\definecolor{myblue}{RGB}{187,187,255}
\definecolor{mygreen}{RGB}{34,139,34}
\definecolor{myorange}{RGB}{255,165,0}
\definecolor{mypink1}{rgb}{0.858, 0.188, 0.478}
\definecolor{mypink2}{rgb}{1, 0.188, 0.478}
\usepackage[colorlinks,bookmarks=false,citecolor=blue,linkcolor=red,urlcolor=blue]{hyperref}

\usepackage{lipsum} 
\usepackage{graphicx}% Include figure files
\usepackage{bm,bbm,bbold}% bold math

\newcommand{\Z}{\mathbb{Z}}
\newcommand{\id}{\mathbb{1}}

\newcommand{\bellstate}{\ensuremath{B^+}}

\usepackage{mathtools}

\newcommand{\CP}{\ensuremath{\textsf{CP}}}
\newcommand{\SWAP}{\ensuremath{\textsf{SWAP}}}

\begin{document}

\title{Tri-unitary quantum circuits}% Force line breaks with \\

\author{Cheryne Jonay}
\affiliation{Department of Physics, Stanford University, Stanford, CA 94305}

\author{Vedika Khemani}
\affiliation{Department of Physics, Stanford University, Stanford, CA 94305}

\author{Matteo Ippoliti}
\affiliation{Department of Physics, Stanford University, Stanford, CA 94305}
\date{\today}%

\begin{abstract}
We introduce a novel class of quantum circuits that are unitary along three distinct ``arrows of time''.
These dynamics share some of the analytical tractability of ``dual-unitary'' circuits, while exhibiting distinctive and richer phenomenology.
We find that two-point correlations in these dynamics are strictly confined to three directions in $(1+1)$-dimensional spacetime -- the two light cone edges, $\delta x=\pm v\delta t$, and the static worldline $\delta x=0$.
Along these directions, correlation functions are obtained exactly in terms of quantum channels built from the individual gates that make up the circuit.
We prove that, for a class of initial states, entanglement grows at the maximum allowed speed up to an entropy density of at least one \emph{half} of the thermal value, at which point it becomes model-dependent.
Finally, we extend our circuit construction to $2+1$ dimensions, where two-point correlation functions are confined to the one-dimensional edges of a tetrahedral light cone -- a subdimensional propagation of information reminiscent of ``fractonic'' physics.
\end{abstract}

\maketitle

\section{Introduction \label{sec:intro}}

The dynamics of quantum many-body systems play a central role in many areas of physics, from non-equilibrium statistical mechanics to applied quantum information science.
Spatiotemporal correlations of local operators are among the most useful physical characterizations of such systems: they diagnose the approach to equilibrium\cite{DAlessio2016} (or lack thereof\cite{Nandkishore2015}), encode transport coefficients\cite{Bertini2021_RMP}, and are more readily measurable in experiments than global properties such as entanglement. 
However, they are notoriously hard to compute for general interacting many-body systems. 

Until recently, spatiotemporal correlations could be calculated exactly only in the presence of integrability~\cite{Bertini2021_RMP, Alba2021}, where the evolution is constrained by extensively many conservation laws.
On the opposite end of the spectrum, much analytical progress on entanglement, thermalization, quantum information scrambling, quantum chaos, and the emergence of hydrodynamics has been achieved via random circuit models\cite{Nahum_2017, vonKeyserlingk_2018, Khemani_2018_Lyapunov, Bertini_2018sff, Nahum_2018, Khemani_2018,  Rakovszky_2018, Chandran_2015, brown2013scrambling, Hayden_2007, Chan_2018, Chalker2020MBL}.
The usefulness of random unitary circuits stems from the fact that they retain only the absolutely fundamental features of quantum dynamics: unitarity and spatial locality. 
Additional structure can then be reintroduced in controlled ways, for instance through symmetries and conservation laws\cite{Khemani_2018, Rakovszky_2018,  Friedman_2019}, or temporal periodicity and the associated set of (Floquet) eigenvalues and eigenvectors.
However, random circuit models are (by design) better suited to the study of ``locally averaged'' quantities, such as entanglement, than of spatiotemporally-resolved ones such as correlators.

\begin{figure}
\centering 
\includegraphics[width=\columnwidth]{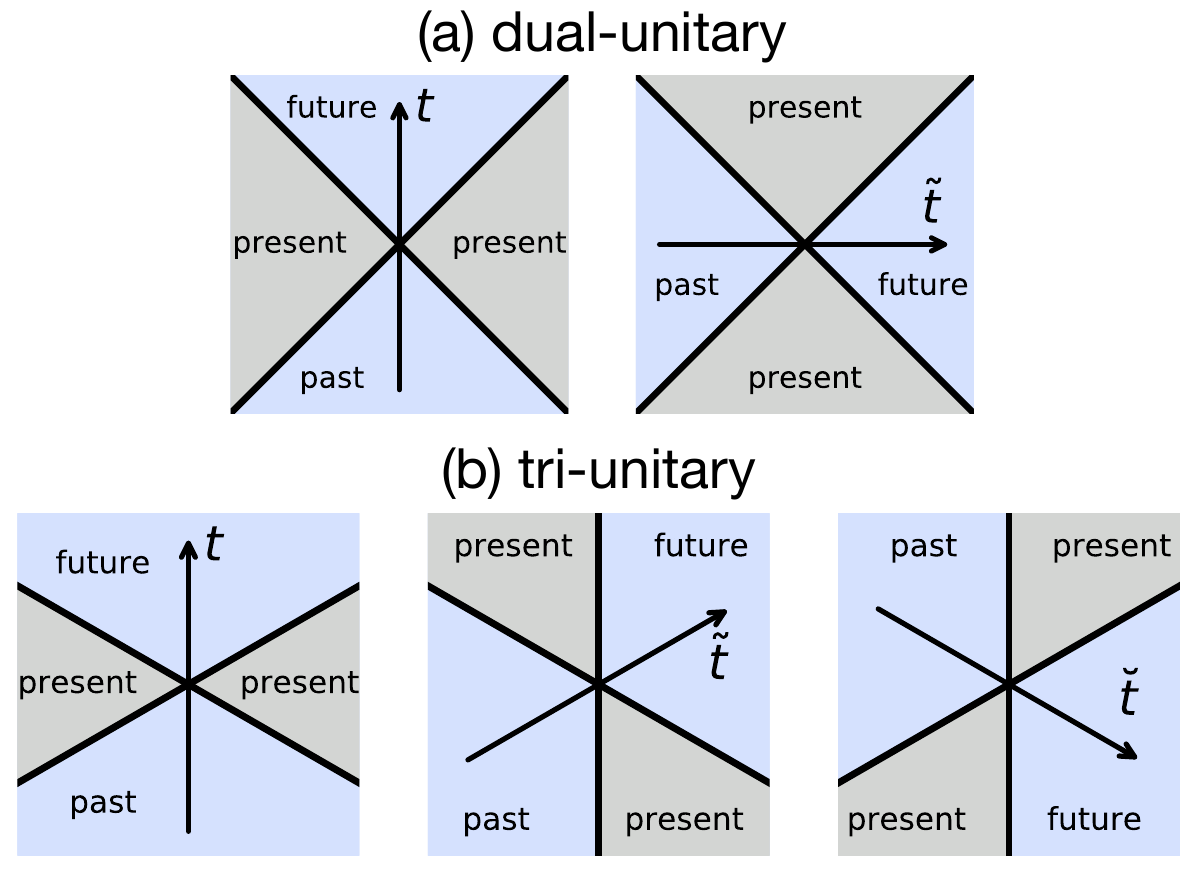}
\caption{Causality structure of dual- and tri-unitary circuits.
(a) Dual-unitary circuits have strict causal light cones under two orthogonal arrows of time, $t$ (left) and $\tilde{t}$ (right). As a result, correlations vanish everywhere except on the light rays $\delta x = \pm \delta t$.
(b) Tri-unitary circuits have strict causal light cones under three arrows of time, $t$ (left), $\tilde{t}$ (center) and $\breve{t}$ (right) at $2\pi/3$ angles with each other. As a result, correlations vanish everywhere except on the intersection of all causally-allowed regions: the light rays $\delta x=\pm v\delta t$ \emph{and} the static worldline $\delta x=0$.} 
\label{fig:overlapping_lightcones}
\end{figure}

It is only recently that a class of local, interacting circuit models was proposed where correlation functions can be calculated analytically\cite{Bertini_2019correlations, Gopalakrishnan_2019}. The key feature of those models is that the dynamics are unitary in the \emph{space} direction, as well as the time direction, as sketched in Fig.~\ref{fig:overlapping_lightcones}(a). 
Such ``dual-unitary'' models have been studied extensively in recent years\cite{Akila_2016, Bertini_2018sff, Bertini_2019entanglement, Bertini_2019correlations, Gopalakrishnan_2019, Bertini_2020SciPost1, Bertini_2020SciPost2, Kos_2020correlations, Piroli_2020exact, gutkin2020local, Claeys_2020, Claeys_2021, Flack2020, Bertini_2020sff, Reid2021, Prosen2021eth, Suzuki2021}, leading to a plethora of exact results on key aspects of quantum many-body dynamics.
In particular, the study of correlations is drastically simplified due to the peculiar causal structure of these models, which only allows correlations between points at ``light-like'' separations\cite{Gopalakrishnan_2019, Bertini_2019correlations}, as can be seen from Fig.~\ref{fig:overlapping_lightcones}(a).

In this work, we introduce a new class of minimal circuit models which we call \emph{``tri-unitary''}. 
They are a family of lattice models in one and two spatial dimensions for which the dynamics are unitary along \emph{three} space-time directions, as sketched (for the one-dimensional case) in Fig.~\ref{fig:overlapping_lightcones}(b).
These models allow us to extend and generalize results obtained on ``dual-unitary'' models.
The phenomenology we uncover is richer in several ways: in one dimension, correlations are pinned to three, not two, possible directions in spacetime; while two of these directions are the edges of a light cone, i.e. ``maximum-velocity'' worldlines just like in the dual-unitary case, the third one is the \emph{static} worldline, $\delta x = 0$. This means that it is possible for information to remain ``stuck'' in place in tri-unitary circuits.
In two dimensions, the phenomenology is even more distinctive: correlations are pinned to three rays in $(2+1)$-dimensional spacetime, namely they propagate at maximum velocity along three special directions in two-dimensional space, while vanishing everywhere else.

Tri-unitary circuits have the property of remaining unitary under six-fold rotations of $(1+1)$-dimensional spacetime. Rather than swapping space and time (``spacetime duality'')~\cite{Ippoliti_2021postselection, Ippoliti_2021fractal, Grover2021, Honeywell2020, Honeywell2021du}, these rotations mix the two nontrivially.
The ensuing dynamics is thus even more ``agnostic'' about the roles of space and time.
This points to interesting connections with recent ideas about the emergence of spacetime from tensor networks~\cite{Swingle2018, Hayden2016, Cotler_2019}, and with holographic quantum error correcting codes -- tensor network architectures that have been recently employed as toy models of the AdS/CFT correspondence\cite{Pastawski_2015}. In both tri-unitary circuits and holographic codes, one has a highly isotropic tensor network that is agnostic about the role of space and time.
The key distinction between the two is that, while holographic codes employ ``perfect tensors'' that always hide information in maximally non-local degrees of freedom, the elementary gates in our tri-unitary circuits allow for information to remain localized and accessible, but only along certain directions.

More concretely, the advent of digital quantum simulators is making it possible to engineer time evolutions that realize target tensor networks by specific sequences of unitary gates and, potentially, projective measurements. 
This flexibility allows one to explore a wide range of unitary\cite{Google2019,Google2021,Ippoliti2020NISQ} and non-unitary\cite{Skinner2019, Li2018, Gullans2020PRX, Honeywell2020, Honeywell2021tn, Noel2021} dynamics, including ones obtained by rotating the ``arrow of time''~\cite{Ippoliti_2021postselection, Ippoliti_2021fractal, Grover2021, Gopalakrishnan_2019, Bertini_2019correlations, Honeywell2021du}. 
Our work thus adds a new direction to this program, by broadening the possibilities for causal structure in tensor networks that are realizable dynamically. 
We note that tri-unitary circuits have the maximal number of unitary ``arrows of time'' that can be embedded in flat $(1+1)$-dimensional spacetime, and thus represent in a sense the ``most isotropic'' architecture for tensor networks that are realizable dynamically in one spatial dimension. 
In addition, $(2+1)$-dimensional tri-unitary circuits extend this program to three-dimensional tensor networks, a much less explored area.

The paper is structured as follows.
In Sec.~\ref{sec:du} we review the definition and properties of dual-unitary circuits, which will serve as a starting point for the following discussion. We introduce tri-unitary circuits (in one spatial dimension) in Sec.~\ref{sec:tu}, and analytically derive their correlation functions in Sec.~\ref{sec:correlations}.
We then prove an exact result about entanglement growth in these circuits in Sec.~\ref{sec:entanglement}, and extend our construction to two spatial dimensions in Sec.~\ref{sec:higher_d}.
We conclude by summarizing our results and pointing to directions for future research in Sec.~\ref{sec:conclusion}.

%%%%%%%%%
% DU
%%%%%%%%%

\section{\label{sec:du}Review of  dual-unitary dynamics}

\begin{figure}
    \centering
    \includegraphics[width=\columnwidth]{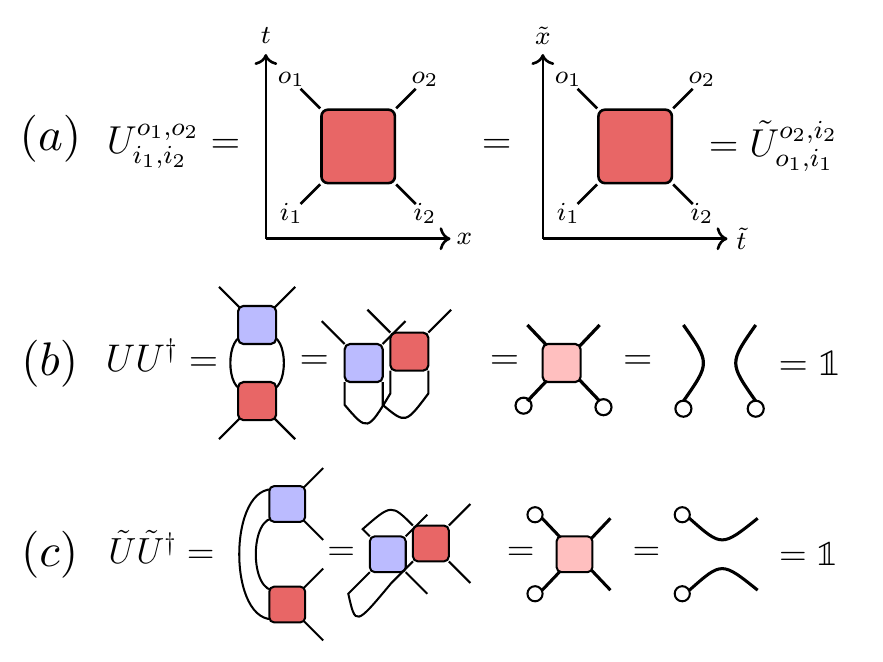}
    \caption{Dual unitarity is a spacetime duality given by the index reshuffling depicted in a). In b) (c)) we show how the contraction of $U$ ($\tilde{U})$ with its Hermitian conjugate simplifies to an identity in the folded picture.}
    \label{fig:dualunitary_identities}
\end{figure}

We begin this review section by introducing {dual-unitary gates}, which serve as the elementary building blocks of dual-unitary circuits. These are two-qubit unitary gates with the special property of being unitary when read ``sideways'', i.e. as a space evolution.
More precisely, we can associate a ``spacetime dual'' gate $\tilde{U}$ to each gate $U$ via a reshuffling of indices,
$\tilde{U}_{i_1 i_2}^{o_1 o_2} = U_{i_1o_1}^{i_2o_2}$, see Fig.~\ref{fig:dualunitary_identities}(a).
Then, dual-unitary gates must satisfy $UU^{\dagger}=U^\dagger U = \id,$ as well as $\tilde{U} \tilde{U}^{\dagger} = \tilde{U}^{\dagger} \tilde{U}=\id$. Those identities are depicted pictorially in Fig.\ref{fig:dualunitary_identities}(b-c). As in Ref.~[\onlinecite{Bertini_2019correlations}], we use the ``folded'' picture, where the unitary $U$ is overlaid with its adjoint $U^{\dagger}$; we also denote contraction with an identity matrix by an open circle $\circ$ terminating a folded tensor's leg.

Arranging dual-unitary gates in a brickwork pattern in $(1+1)$-dimensional spacetime yields a dual-unitary circuit. 
These circuits realize examples of ``maximally chaotic'' dynamics and allow considerable analytical tractability: exact results have been obtained for their spectral form factor\cite{Bertini_2018sff, Flack2020, Bertini_2020sff}, state\cite{Bertini_2019entanglement, Piroli_2020exact} and operator\cite{Bertini_2020SciPost1, Bertini_2020SciPost2, Reid2021} entanglement growth, two-point correlation functions\cite{Gopalakrishnan_2019, Bertini_2019correlations, Kos_2020correlations, gutkin2020local} and out-of-time-ordered correlators\cite{Claeys_2020}.
A remarkable physical property of these circuits, underlying several of the above results, stems from their causal structure.
Conventionally, unitarity and the strict geometric locality of the circuit structure forbid correlations between points displaced by a spacetime vector $(x,t)$ if $|x|>|t|$, i.e. if the two points lie outside each other's (past or future) light cone. 
By the same token however, in dual-unitary circuits one can view the evolution ``sideways'' and rule out correlations if $|t|>|x|$ (in a sufficiently large system).
As a result, correlations are only possible on the light cone's boundary, $|x|=|t|$. So any correlations must propagate exactly at the speed of light. 

To set up the discussion in the following sections, it is helpful to review the behavior of two-point functions in dual-unitary circuits in greater detail.
Infinite-temperature two-point functions obey\cite{Gopalakrishnan_2019, Bertini_2019correlations}
\begin{align}
    C^{ab}(x,t) & = \frac{1}{2^L} \Tr (a_0(t) b_x(0)) \nonumber \\
    & = \frac{1}{2^L} \sum_{\mu =\pm 1} \delta_{x,\mu t} \Tr(M_\mu^{t} (a_0) b_x)\;,
\end{align}
where the $M_{\pm}$ are ``transfer matrices'' (quantum channels) which move an operator along either the left or right fronts of the light cone, and are obtained directly from the gates $U$ that make up the circuits\footnote{Namely, we have $M_+ (a) \equiv \Tr_1( U^\dagger a_1 U)$ and $M_-(a) \equiv \Tr_2 (U^\dagger a_2 U)$ where $U$ acts on qubits 1, 2 and $a_1\equiv a\otimes\id$, $a_2\equiv \id\otimes a$.}. (Here one unit of $t$ is defined as half of a brickwork layer.)
Correlations are then obtained analytically by diagonalizing the single-qubit quantum channels $M_{\pm}$.
Since these are trace-preserving and unital (i.e. $M_\pm(\id) = \id$), one only has to consider the traceless subspace spanned by $X$, $Y$ and $Z$ Pauli matrices;
denoting the three eigenvalues in this subspace by $\{\lambda_{\pm,i}:i=1,2,3\}$, for traceless operators $a,b$, we have
\begin{align}
    C^{ab}(x,t) &=\sum_{\mu = \pm 1} \delta_{x,\mu t} \sum_{i=1,2,3} c^{ab}_{\mu,i} \lambda_{\mu, i}^{t}\;,
\end{align}
where the coefficients $c^{ab}$ are overlaps between the operators $a,b$ and the eigenmodes of $M_\pm$. 
This places strong constraints on the time dependence of correlators.
In particular, depending on the eigenvalues $\lambda_{\mu,i}$, one has the following possible types of behavior\cite{Bertini_2019correlations, Claeys_2021}: 
(i) non-interacting (all eigenvalues are 1, correlations are constant on the light cone); 
(ii) interacting non-ergodic (some but not all eigenvalues are 1, some correlations are constant while others decay); 
(iii) ergodic non-mixing (none of the eigenvalues are 1, but there is at least one unimodular eigenvalue, some correlations oscillate indefinitely); 
and (iv) ergodic and mixing (all $|\lambda_{\mu,i}|<1$, all correlations decay exponentially).

We conclude this review by recalling the general parametrization of dual-unitary gates on two qubits, 
\begin{equation}
    U_{\text{d.u.}}[\phi ] \equiv u_1 u_2 \cdot \SWAP_{1,2} \cdot \CP_{1,2} (\phi) \cdot v_1 v_2 \;,
    \label{eq:du_parametrization}
\end{equation}
where $u_{1,2}$, $v_{1,2} \in SU(2)$ are arbitrary single-qubit gates (tensor product symbols are omitted) and $\CP(\phi) \equiv e^{-i \frac{\phi}{4} (Z_1-\id)(Z_2-\id)}$ is a controlled-phase gate (despite the slightly different notation, the parametrization in Eq.~\eqref{eq:du_parametrization} is equivalent to the one in Ref.~[\onlinecite{Bertini_2019correlations}]).

%%%%%%%%%
% TU
%%%%%%%%%

\section{\label{sec:tu}Tri-unitarity}

As we have seen, the usefulness of dual-unitary circuits for studying quantum dynamics largely stems from their unusual causal structure: having two equally valid ``arrows of time'', they remain unitary under four-fold rotations of spacetime. 
It is thus interesting to generalize this idea and expand the domain of nontrivial many-body quantum dynamics amenable to analytical treatment via rotations in spacetime.
To this end, we introduce a new class of ``tri-unitary'' circuits, which admit \emph{three} unitary arrows of time and remain unitary under {six-fold} rotations of $(1+1)$-dimensional spacetime. 
In this Section we first present the architecture of such circuits in $(1+1)$ dimensions, then the tri-unitarity condition on individual gates, and finally a realization of a family of tri-unitary circuits as time-dependent local Hamiltonians of the kicked Ising type.

\subsection{\label{sec:tu_circuits}Tri-unitary circuits}

\begin{figure}
\includegraphics[width=\columnwidth]{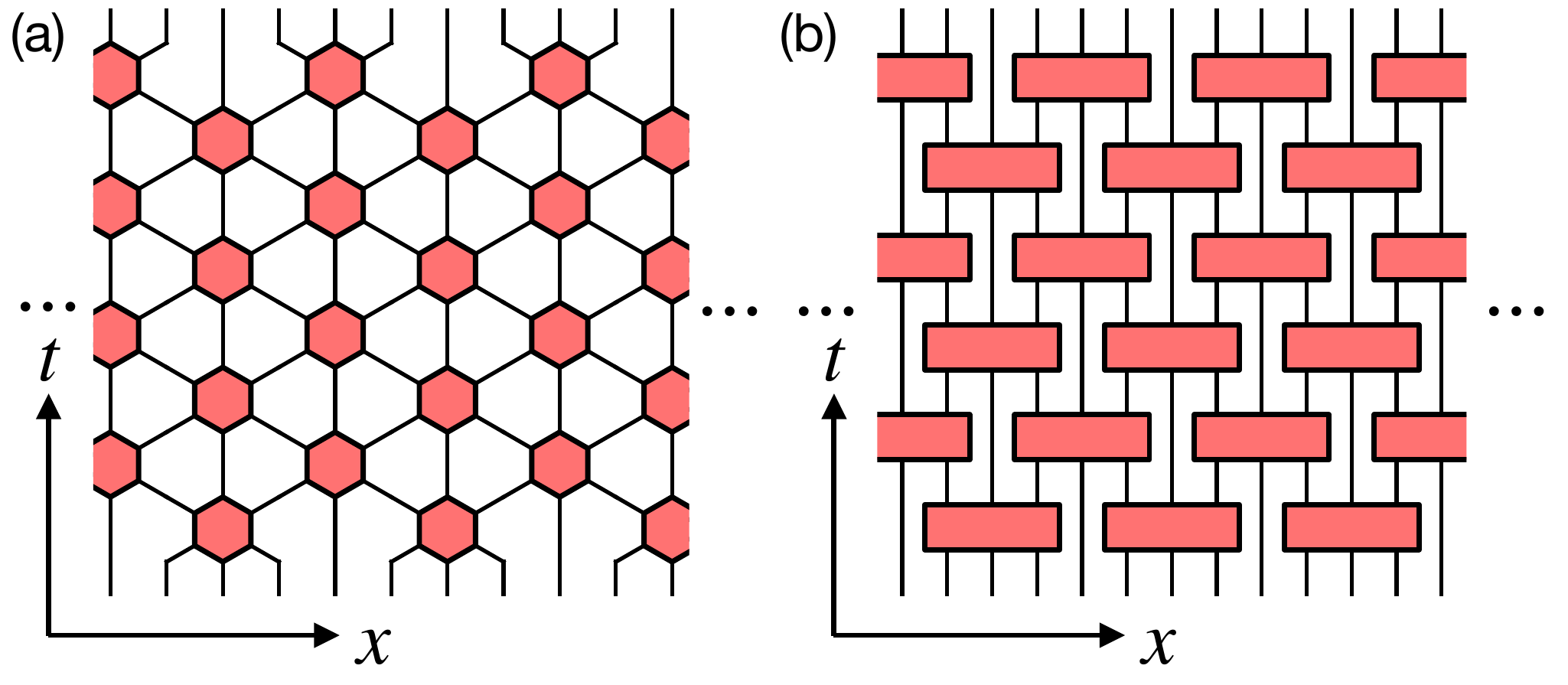}
\caption{Tri-unitary circuit layout. 
(a) Tri-untiary gates (hexagons) are arranged at the sites of a triangular lattice in spacetime, with manifest six-fold rotational symmetry. 
(b) Same circuit drawn with rectangular gates and vertical qubit worldlines.
}
\label{fig:triunitary_circuits}
\end{figure}

Much like dual-unitary circuits are built from dual-unitary gates arranged on a square lattice in spacetime, tri-unitary circuits require their own special set of gates arranged on a \emph{triangular} lattice in spacetime.
This is shown in Fig.~\ref{fig:triunitary_circuits}(a), where the hexagons represent three-qubit gates (constraints on these gates are discussed in the following). 
We note that this construction does not extend beyond three arrows of time, at least in flat Minkowski spacetime: there are no Bravais lattices with coordination higher than 6, and thus no natural ways of building ``$n$-unitary circuits'' for $n>3$. 

Additionally, Fig.~\ref{fig:triunitary_circuits}(b) shows the same circuit architecture with more familiar rectangular gates and vertical qubit worldlines, clarifying the layout for practical implementations.
Notice there is a spatial unit cell of 4 qubits, with an important difference between even qubits (which take part in all the interactions) and odd ones (which skip every other layer).
Formally, a brick-work layer of the resulting unitary circuit is given by $\mathbb{U}_o \mathbb{U}_e$, with
\begin{align}
    \mathbb{U}_e &= \bigotimes_{x \in 4\Z} (U_{x,x+1,x+2} \otimes \id_{x+3}), \>\> \mathbb{U}_o = \mathbb{T}^2 \mathbb{U}_e (\mathbb{T}^{\dagger})^2,
    \label{eq:brickwork}
\end{align}
where $U$ is a three-qubit gate and $\mathbb{T}$ is the one-site translation operator.
Notice that in each layer, one out of every 4 qubits is not acted upon.

The architecture in Fig.~\ref{fig:triunitary_circuits}(a) is manifestly invariant under six-fold rotations of spacetime, however the circuit resulting from such rotation is, in general, not unitary. For this to be the case, we must constrain the choice of three-qubit gates $U$.

\subsection{\label{sec:tu_gates}Tri-unitary gates}

\begin{figure}
    \centering
    \includegraphics[width=0.9\columnwidth]{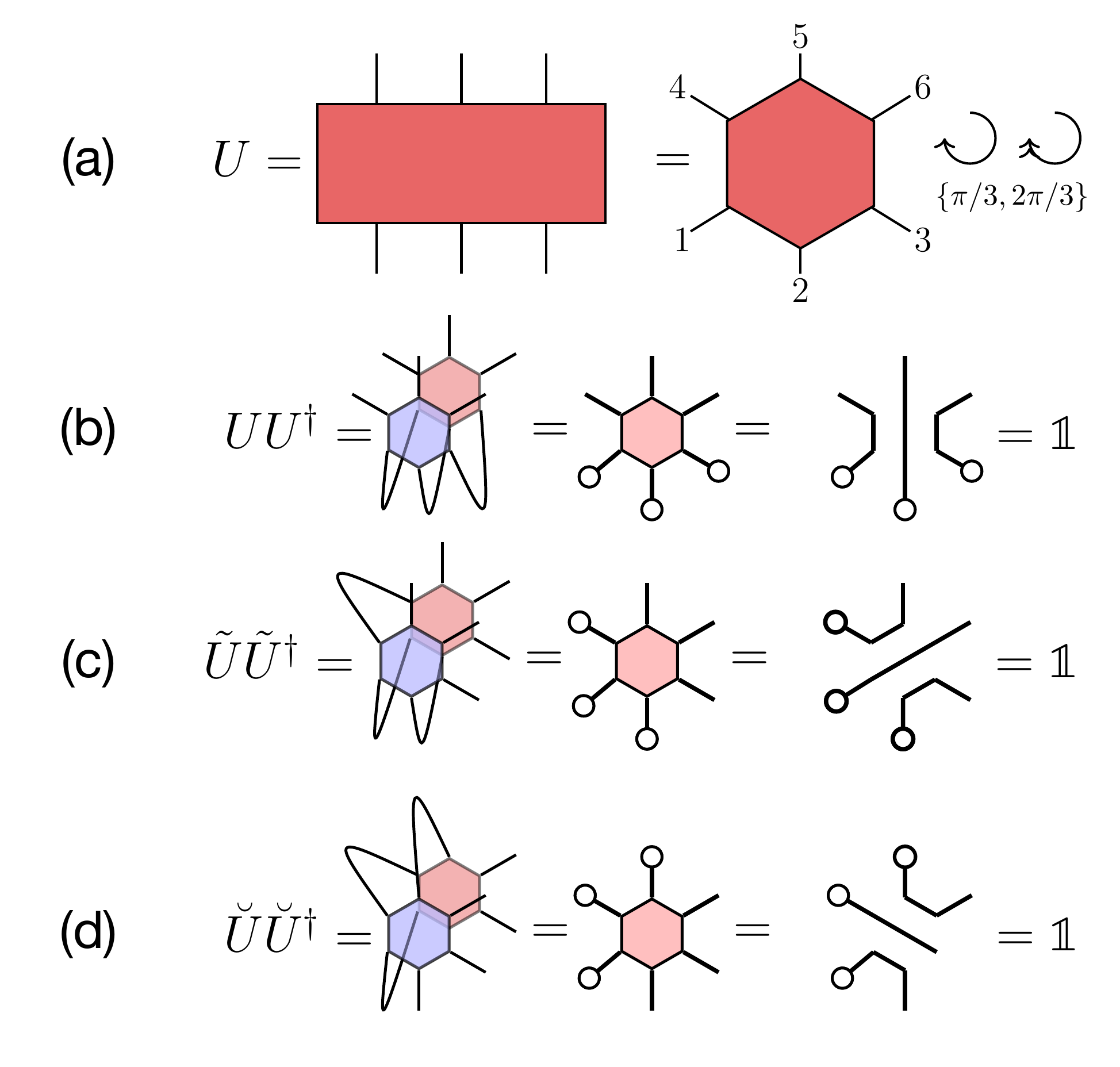}
    \caption{
    (a) A three-qubit gate $U$, normally drawn as a rectangle (left), can be represented as a hexagon in order to facilitate spacetime rotations (right).
    Legs are labeled from $1$ to $6$.
    (b) Unitarity of $U$. The blue hexagon represents $U^\ast$; the pink hexagon represents $U\otimes U^\ast$ in the ``folded picture'', where each leg is implicitly doubled. Open circles represent contraction with $\id$ operators. 
    (c, d) Unitarity of $\tilde{U}$ and $\breve{U}$: picking different triplets of (contiguous) input legs still yields a unitary operator. }
    \label{fig:triunitary_identities}
\end{figure}

We denote the input legs as $(1,2,3)$ and the output legs by $(4,5,6)$, as shown in Fig.~\ref{fig:triunitary_identities}(a).
The requirement of tri-unitarity is that these tensors be unitary under three distinct choices for the arrow of time, i.e., whether we map qubits $(1,2,3)\mapsto (4,5,6)$ (the ``conventional'' arrow of time, Fig.\ref{fig:triunitary_identities}(b)), $(4,1,2)\mapsto (5,6,3)$ (clockwise $\pi/3$ rotation, Fig.\ref{fig:triunitary_identities}(c)), or $(5,4,1) \mapsto (6,3,2)$ (clockwise $2\pi/3$ rotation, Fig.\ref{fig:triunitary_identities}(d)).
This selects three distinct contractions of $U$ and $U^\ast$ and sets them equal the three-qubit identity operator. 
More specifically, we can define transformed gates $\tilde{U}$ and $\breve{U}$ via
\begin{equation}
U^{a_4 a_5 a_6}_{a_1 a_2 a_3} =  \tilde{U}^{a_5 a_6 a_3}_{a_4 a_1 a_2}  =  \breve{U}^{a_6 a_3 a_2}_{a_5 a_4 a_1}  \;.
\label{eq:tu_def}
\end{equation}
(notice the counterclockwise rotation of indices).
Then, a gate $U$ is tri-unitary if it satisfies all three identities\footnote{We note that tensors obeying these same constraints (for generic numbers of legs, thus also including dual-unitarity) were introduced in Ref.~[\onlinecite{Harris2018}] as ``block-perfect tensors'' in the context of holographic quantum error correcting codes, in Ref.~[\onlinecite{Berger2018}] as ``perfect tangles'' for modular tensor categories, and in in Ref.~[\onlinecite{Doroudiani2020}] as ``planar maximally entangled states''.}
 $UU^\dagger = \tilde{U}\tilde{U}^\dagger = \breve{U}\breve{U}^\dagger = \id$, shown in Fig.~\ref{fig:triunitary_identities}(b-d).

First of all, it is natural to ask whether the conditions in Eq.~\eqref{eq:tu_def} can be satisfied at all. 
The answer is positive; in fact, one can do even more: it is known that the 5-qubit perfect quantum error-correcting code\cite{Bennett1996, Laflamme1996} can be used to construct a ``perfect tensor'', i.e. a 6-qubit tensor such that \emph{any} bipartition of its legs into 3 inputs and 3 outputs (not necessarily contiguous) yields a unitary gate\cite{Pastawski_2015}. 
There are 10 such partitions;
by contrast, tri-unitarity only constraints three bipartitions. 
Perfect tensors (also known as ``absolutely maximally entangled states''~\cite{Helwig2012, Goyeneche2015, Linowski2020}) play an important role in toy models of holography\cite{Pastawski_2015, Latorre2015, Hayden2016}, where such highly isotropic tensors are arranged on a lattice in curved space; this results in a quantum error correcting code that encodes bulk spacetime degrees of freedom into the edge, in a manner reminiscent of the AdS/CFT correspondence. 
While perfect tensors can be naturally viewed as quantum error-correcting codes, generic tri-unitary gates can be viewed as ``defective'' quantum codes: the encoded qubit is vulnerable to errors on a specific physical qubit. Thus, crucially, some of the encoded information can be accessed locally, but only along a special direction.
As we shall see in Sec.~\ref{sec:correlations}, this geometrically constrained leakage of information has a striking effect on the behavior of two-point correlations in generic tri-unitary circuits.  

\begin{figure}
\centering 
\includegraphics[width=\columnwidth]{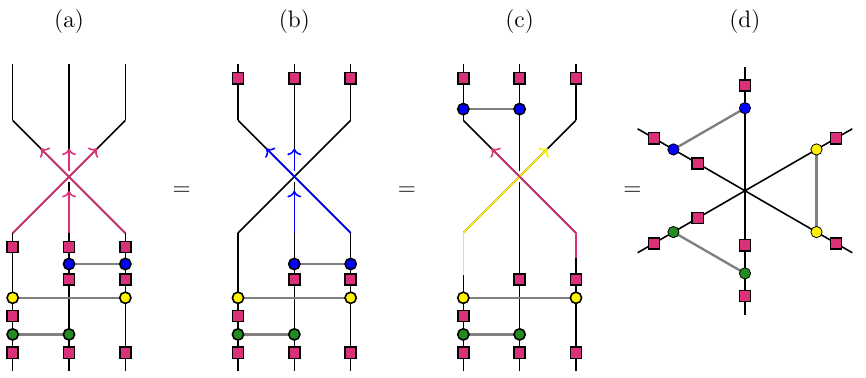}
\caption{(a) Diagrammatic representation of the gate $U_\text{t.u.}[\boldsymbol{\phi}]$ in Eq.~\eqref{eq:UCP}, consisting of two-qubit $\CP$ gates (dots joined by horizontal lines), single-qubit rotations (squares), and a $\SWAP$ gate (exchange of left and right qubit worldlines). 
(b,c)
This diagram can be recast in a manifestly tri-unitary form by ``sliding'' one- and two-qubit gates past the $\SWAP$.
The last step (sliding one end of the yellow $\CP$ gate past the $\SWAP$) relies on dual-unitarity of $\SWAP\cdot \CP$.
(d) The resulting diagram is manifestly invariant under rotations of spacetime by $2\pi/3$ (up to a permutation of parameters), thus the gate is tri-unitary.
}
\label{fig:CP_gate}
\end{figure}

While a complete characterization of all tri-unitary gates is a goal for future work, here we will focus on a particularly simple family of tri-unitary gates\footnote{A similar construction of ``planar maximally entangled states'' in Ref.~[\onlinecite{Doroudiani2020}] yields a subset of this family.}, which are simple to construct and sufficient to realize all the relevant physics.
These gates are constructed out of controlled-phase gates $\CP_{ij}(\phi) \equiv e^{-i\frac{\phi}{4}(Z_i-\id)(Z_j-\id)}$ between pairs of qubits followed by a $\SWAP$ gate between qubits 1 and 3, in analogy to the dual-unitary gate parametrization in Eq.~\eqref{eq:du_parametrization}:
 \begin{align}
     U_\text{t.u.}[\boldsymbol{\phi}] & = u_1 u_2 u_3 \cdot \SWAP_{3,1} \cdot \CP_{3,1} (\phi_3)  \cdot v_1 v_3  \nonumber \\ & \qquad \cdot \CP_{2,3}(\phi_2) \cdot v_2  \cdot \CP_{1,2}(\phi_1) \cdot w_1 w_2 w_3 \;,
     \label{eq:UCP}
 \end{align}
where $u_i$, $v_i$, $w_i$ are arbitrary single-qubit gates acting on qubit $i$ (tensor product symbols are omitted). 
This family of gates is sketched in Fig.~\ref{fig:CP_gate}(a).
Their tri-unitarity is made apparent by drawing them in a symmetric way, as in Fig.~\ref{fig:CP_gate}(d): a $2\pi/3$ rotation results in a permutation of the single-qubit gates and the $\phi$ angles, preserving the family of gates as a whole; rotations by $\pi/3$ on the other hand produce the \emph{transpose} of a gate in the family, which is still unitary. Thus $U$, $\tilde{U}$ and $\breve{U}$ are all unitary, making $U$ tri-unitary.
Intuitively, one obtains Fig.~\ref{fig:CP_gate}(d) by ``sliding'' the controlled-phase gates past the \textsf{SWAP} in Fig.~\ref{fig:CP_gate}(a), following the steps sketched in Fig~\ref{fig:CP_gate}(b,c).
More rigorously, we note that $\SWAP_{1,3} \CP_{1,3}(\phi_3)$ equals the dual-unitary gate $U_{\text{d.u.}}[\phi_3]$ of Eq.~\eqref{eq:du_parametrization} acting on qubits 1 and 3; thus the vertical (yellow) $\CP$ gate in Fig.~\ref{fig:CP_gate}(d), which at first glance appears ill-defined (coupling the same qubit at different times), is in fact equivalent via spacetime duality to a unitary gate acting (simultaneously) on two distinct qubits.

This family of gates is specified by 31 parameters (nine $SU(2)$ gates, with three angles each, the three $\boldsymbol{\phi}$ interaction phases, and a global phase).
While not a full parametrization of tri-unitary gates (e.g., it does not contain the perfect tensor\footnote{The gates defined in Eq.~\eqref{eq:UCP} have at most two bits of entropy between non-contiguous bipartitions of the type $(1,2,6) \mapsto (3,4,5)$. As such, they cannot express the perfect tensor, whose entropy across any bipartition of the 6 legs is maximal (3 bits). 
}), this is nonetheless a very large space that offers a rich set of possibilities for dynamics.

\subsection{\label{sec:ki} Kicked Ising model realization}

\begin{figure}
    \centering
    \includegraphics[width=\columnwidth]{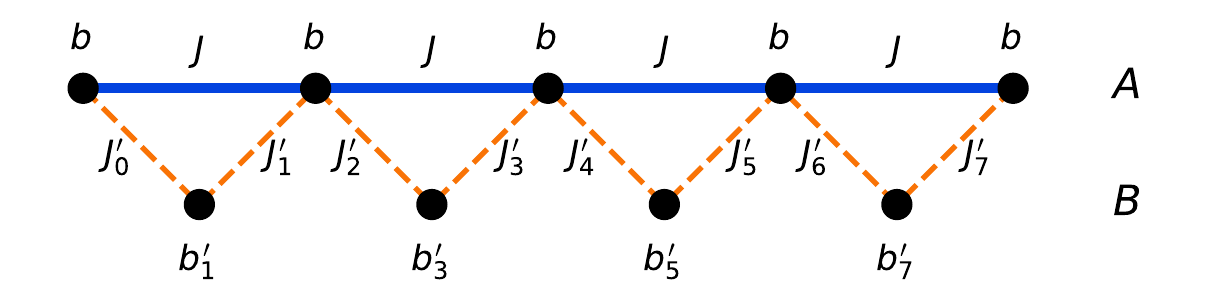}
    \caption{Schematic of the kicked Ising model realization of tri-unitary dynamics. 
    Circles represent qubits, partitioned into two sublattices $A$ and $B$; the $J$ and $J'$ bonds denote $ZZ$ Ising interactions; $b$ and $b'$ are on-site transverse fields.
    For $|J| = |b| = \pi/4$ (and arbitrary $b'$, $J'$) the Floquet operator Eq.~\eqref{eq:KI_TU} is tri-unitary.}
    \label{fig:ki}
\end{figure}

While the tri-unitary circuits constructed above would most naturally be realized on gate-based digital quantum simulators, they may also be realized as time-dependent Hamiltonians.
As an example, we consider a spin chain partitioned in two sublattices, $A$ and $B$, evolving under a generalized kicked-Ising dynamics as follows.
The spin chain's Hamiltonian alternates between three forms:
a transverse field, $H_X = b\sum_{n\in A} X_n + \sum_{n\in B} b'_n X_n$; 
a next-nearest neighbor Ising coupling in the $A$ sublattice, $H_Z^{AA} = J \sum_{n\in A} Z_{n} Z_{n+2}$;
and the same interaction $H_Z^{AA}$ with an additional nearest-neighbor Ising coupling (coupling sublattices $A$ and $B$), $H_Z^{AB} = \sum_n J'_n Z_n Z_{n+1}$. 
The setup is summarized schematically in Fig.~\ref{fig:ki}.
(In addition one could include arbitrary longitudinal fields $h_nZ_n$ in the $H_Z$ Hamiltonians.) 
The Floquet unitary is given by 
\begin{equation}
    U_F = e^{-i H_X} e^{-iH_Z^{AA}} e^{-iH_X} e^{-i (H_Z^{AA} + H_Z^{AB})} \;.
    \label{eq:KI_TU}
\end{equation}
At $J=b=\pi/4$ (and for arbitrary, possibly position-dependent $J'$, $b'$, $h$), the above evolution is tri-unitary.
This is most easily seen by setting $J'=0$ first: in that case, sublattice $A$ realizes a kicked-Ising chain at the dual-unitary point, as in Ref.~[\onlinecite{Bertini_2018sff}], while sublattice $B$ consists of decoupled sites. 
This realizes the tri-unitary gate $U_\text{t.u.}[\boldsymbol{\phi}]$, Eq.~\eqref{eq:UCP}, with $\phi_1=\phi_2=0$ and $\phi_3 = \pi$ (i.e. a dual-unitary gate $U_\text{d.u.}[\pi]$ on sites 1 and 3, while site 2 is untouched). 
Turning on the $J'$ couplings preserves this structure, but also introduces nontrivial $\phi_{1,2}$ interaction angles, coupling the two sublattices.
This allows one to explore nontrivial tri-unitary gates $U_\text{t.u.}[\boldsymbol\phi]$ of arbitrary $\boldsymbol{\phi}$ angles, where the three arrows of time are fully on the same footing.

%%%%%%%%%%%%%%%
% CORRELATIONS
%%%%%%%%%%%%%%%

\section{\label{sec:correlations}Correlations in tri-unitary circuits}

For concreteness we consider spatiotemporally uniform (i.e. clean, Floquet) tri-unitary circuits, built out of a single tri-unitary gate $U$ arranged in spacetime according to Fig.~\ref{fig:triunitary_circuits}(a). 
We consider two-point correlations between two traceless, single-site operators $a$ and $b$, displaced by $x$ sites and $t$ time steps (again one unit of $t$ consists of a half brickwork layer), at infinite temperature:
\begin{equation}
    C^{ab}(x,t) = \frac{1}{2^L} \Tr(a_0(t) b_x(0))\;.
\end{equation}
The usual light cone for time evolution with $U$ forces $C^{ab}$ to vanish if $|x|>v|t|$, with $v=2$ (each three-qubit gate can move information by two sites). 
Normally (for $U$ unitary but not tri-unitary), correlations can exist anywhere inside the light cone;
however, the two additional directions of unitarity ($\tilde{U}$ and $\breve{U}$) pose further restrictions.
As sketched in Fig.~\ref{fig:overlapping_lightcones}, each direction of unitarity rules out correlations in two $\pi/3$ wedges of spacetime that are in the ``present'' relative to the arrow of time. 
The intersection of these three constraints rules out almost all of spacetime, leaving only the lines $x = \pm vt$, $x=0$.

\begin{figure}
\centering 
\includegraphics[width=\columnwidth]{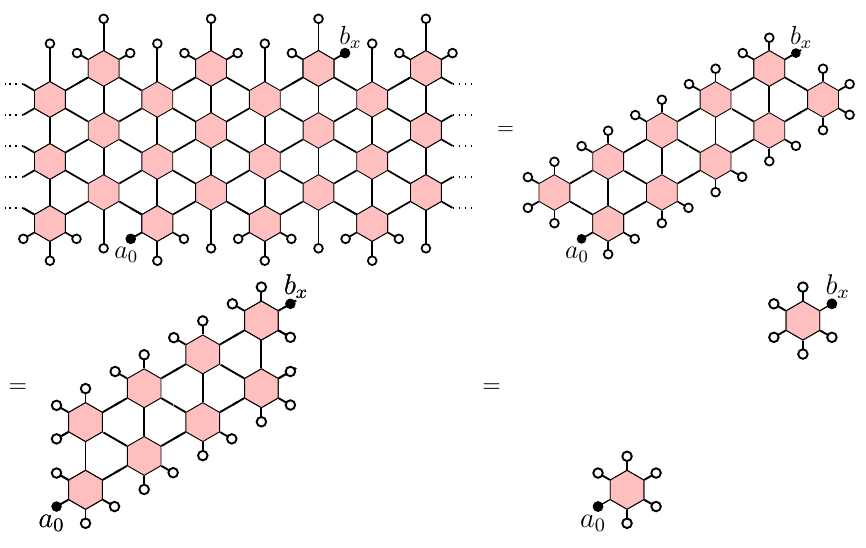}
\caption{Correlation function $\langle a_0(0) b_x(t) \rangle$ for $0<x<vt$ (away from special rays $x=0,\pm vt$). Hexagons are ``folded'' gates $U\otimes U^\ast$, open circles are contractions to $\id$, full circles are the $a$, $b$ operators. 
The first equality follows from unitarity of $U$; the second from unitarity of either $\tilde{U}$ or $\breve{U}$; the last from unitarity of $\breve{U}$. 
Thus the correlator becomes disconnected (and vanishes for traceless $a$, $b$).
In the reflected ($x\mapsto -x$) version of this diagram, the last step would instead require unitarity of $\tilde{U}$. }
\label{fig:correlation_doesnotsurvive}
\end{figure}

More formally, using the tri-unitarity identities $UU^\dagger = \tilde{U}\tilde{U}^\dagger = \breve{U} \breve{U}^\dagger = \id$, in their diagrammatic form, Fig.~\ref{fig:triunitary_identities}(b-d), we can explicitly reduce the tensor contraction expressing $C^{ab}(x,t)$ and show that it vanishes away from these special lines: Fig.~\ref{fig:correlation_doesnotsurvive} shows an example with $0<x<vt$, in which unitarity of $U$ and $\breve{U}$ causes the correlator to vanish. Symmetrically, for $-vt<x<0$, one would need unitarity of $U$ and $\tilde{U}$ to reach the same result.

\begin{figure}
\centering 
\includegraphics[width=\columnwidth]{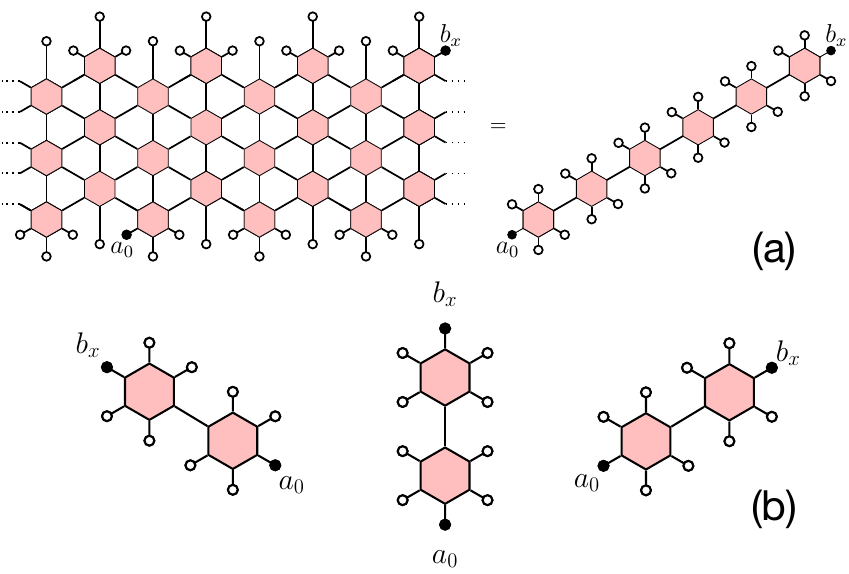}
\caption{(a) Correlation function $\langle a_0(0) b_x(t)\rangle$, for $x=vt$. Unlike Fig.~\ref{fig:correlation_doesnotsurvive}, the correlator remains connected and is given by iteration of a quantum channel, $M_+$. Analogous results would be obtained for $x=-vt$ and $x=0$ (in terms of $M_-$ and $M_0$ respectively).
(b) The quantum channels $M_-$, $M_0$ and $M_+$: the diagrams show $\Tr(b M_\mu^2(a))$ with $\mu = -$ (left), $0$ (center), $+$ (right).}
\label{fig:correlation_survives}
\end{figure}

For $x = 0$ or $x=\pm vt$, we can exactly express the correlations in terms of iterated application of one of three quantum channels, $M_\pm$ (for correlations along $x=\pm vt$) and $M_0$ (for correlations along $x=0$), as shown in Fig.~\ref{fig:correlation_survives}(a). 
This phenomenology is similar to that of dual-unitary circuits, which also exhibit left- and right-moving correlations along $x=\pm vt$; the ``non-moving'' auto-correlator $x=0$, in contrast, is unique to the tri-unitary setting. 

The quantum channels $M_\mu$ ($\mu = 0,\pm$) are given in terms of the tri-unitary gate $U$ as
\begin{align}
M_-(a) &= \frac{1}{4} \Tr_{2,3} [U^{\dagger} (\id \otimes\id\otimes a) U] \label{eq:M-} \\
M_0(a)&= \frac{1}{4} \Tr_{1,3}[ U^{\dagger} (\id \otimes a \otimes \id) U] \label{eq:M0} \\
M_+(a) &= \frac{1}{4} \Tr_{1,2} [U^{\dagger} (a \otimes\id\otimes\id) U] \label{eq:M+}
\end{align}
where $\Tr_{ij}$ is the partial trace over sites $i$ and $j$. 
These maps are quantum channels (up to a $\SWAP$, they correspond to coupling a qubit to two ancillas in an infinite-temperature state, evolving unitarily, and discarding the ancillas).
Furthermore, they are unital, i.e. $M_{\pm,0}(\id) = \id$. 
As a result, the traceless subspace (spanned by the $X$, $Y$, $Z$ Pauli matrices) does not mix with the $\id$ subspace, and all information about correlations is obtained by diagonalizing the $3\times 3$ matrices $M_{\mu}^{\alpha\beta}\equiv \Tr(\sigma_\alpha M_\mu(\sigma_\beta))/2$, where $\mu = \pm,0$ and $\alpha,\beta = x,y,z$.

We can observe all four of the behaviors listed in Sec.~\ref{sec:du} (from non-interacting to ergodic and mixing) within the family of gates in Eq.~\eqref{eq:UCP} and Fig.~\ref{fig:CP_gate}.
The simplest example is given by setting $\phi_i \equiv 0$ and all single-qubit gates to $\id$, giving $U = \SWAP_{1,3}\otimes \id_2$: 
all $M_\mu$ act as identity channels, and thus all correlations along the special rays are constant. 
This is intuitive, as the circuit in this limit reduces to a sequence of $\SWAP$s acting on odd qubits (giving left- and right-movers), while even qubits are inert (``non-movers'').
Letting $\phi_i\neq 0$ makes the dynamics interacting but non-ergodic ($ZZ$ correlators are along the rays while other decay); 
by introducing generic single-qubit gates one realizes all other behaviors, which we survey in more detail in Appendix~\ref{app:CPcorrelations}.

Finally we note that if $U$ is a perfect tensor, then contracting \emph{any} three legs in $U\otimes U^\ast$ with $\id^{\otimes 3}$ necessarily yields $\id^{\otimes 3}$ on the three remaining legs; thus all $M_\mu$ (whose definition involves contraction with $\id^{\otimes 4}$ on four non-contiguous legs) reduce to exact erasure channels, $M_\mu(a) = \Tr(a) \id/2$, and all two-point correlations vanish everywhere.

%%%%%%%%%%%%%%%
% ENTANGLEMENT
%%%%%%%%%%%%%%%

\section{\label{sec:entanglement} Entanglement growth}

Another question that is made analytically tractable by tri-unitarity is the growth of entanglement.
We show here that, starting from a class of simple, short-range-entangled initial states, the half-chain entanglement entropy in the thermodynamic limit grows at the maximum velocity allowed by the circuit's geometry, $S(t) = t$ (in bits). 
For a finite subsystem of $\ell$ qubits, we find $S(t) = 2t$ (each boundary contributing $t$) but only up to a time $t = \ell/4$, at which point the entropy density is $S/\ell = 1/2$; afterwards, entanglement growth becomes model-dependent. 

In this Section, we will allow the local Hilbert space dimension to be $q \geq 2$, and measure entropy in units of $\ln(q)$ (i.e. bits for $q=2$, etc).
As in the case of dual-untiary circuits, it is helpful to consider a suitable class of ``solvable'' initial states\cite{Piroli_2020exact}. While the task of classifying all such states in the tri-unitary case is left for future work, a particularly simple family of such states is given by Bell pairs intercalated by disentangled sites:
\begin{equation}
    | \psi_0 \rangle = \bigotimes_{x \in 4\mathbb Z} |\bellstate \rangle_{x-2,x} |0\rangle_{x-1} |0 \rangle_{x+1} \;,
    \label{eq:solvable_initstate}
\end{equation}
with $|\bellstate\rangle \equiv \frac{1}{\sqrt{q}} \sum_{j=0}^{q-1} |jj\rangle$ a Bell pair state on two qudits, and $|0\rangle$ could be replaced by any single-qudit state (a short-range-entangled state would cause a minor change to the proof).
Notice that the first unitary layer to act on $\ket{\psi_0}$ is $\mathbb{U}_e$ in Eq.~\eqref{eq:brickwork}, with three-qubit gates acting on triplets $(x,x+1,x+2)$,  $x\in 4 \mathbb{Z}$, while the initial entanglement is between qubits $(x-2,x)$, $(x+2, x+4)$, etc.

We consider a semi-infinte contiguous subsystem $A = \{x<x_\text{cut}\}$ in an infinite chain $x\in \mathbb{Z}$.
The initial state in Eq.~\eqref{eq:solvable_initstate} makes the computation of $\Tr \rho_A^n$ analytically tractable, for all integer $n\geq 2$, thus giving all the Renyi entanglement entropies $S_n=1/(1-n) \log \Tr(\rho^n)$, $n\geq 2$. 
We will find that $S_n(t)$ is independent of $n$ and equals either $t$ or $t+1$, depending on the entanglement cut $x_\text{cut}$ and the parity of $t$.  
The fact that $S_n(t)$ is independent of $n$ for all integer $n\geq 2$ implies that all the Renyi entropies, as well as the von Neummann entropy, in fact coincide. This allows us to use the word `entropy' in an unqualified sense in this setting.

\begin{figure}
    \centering
    \includegraphics[width=\columnwidth]{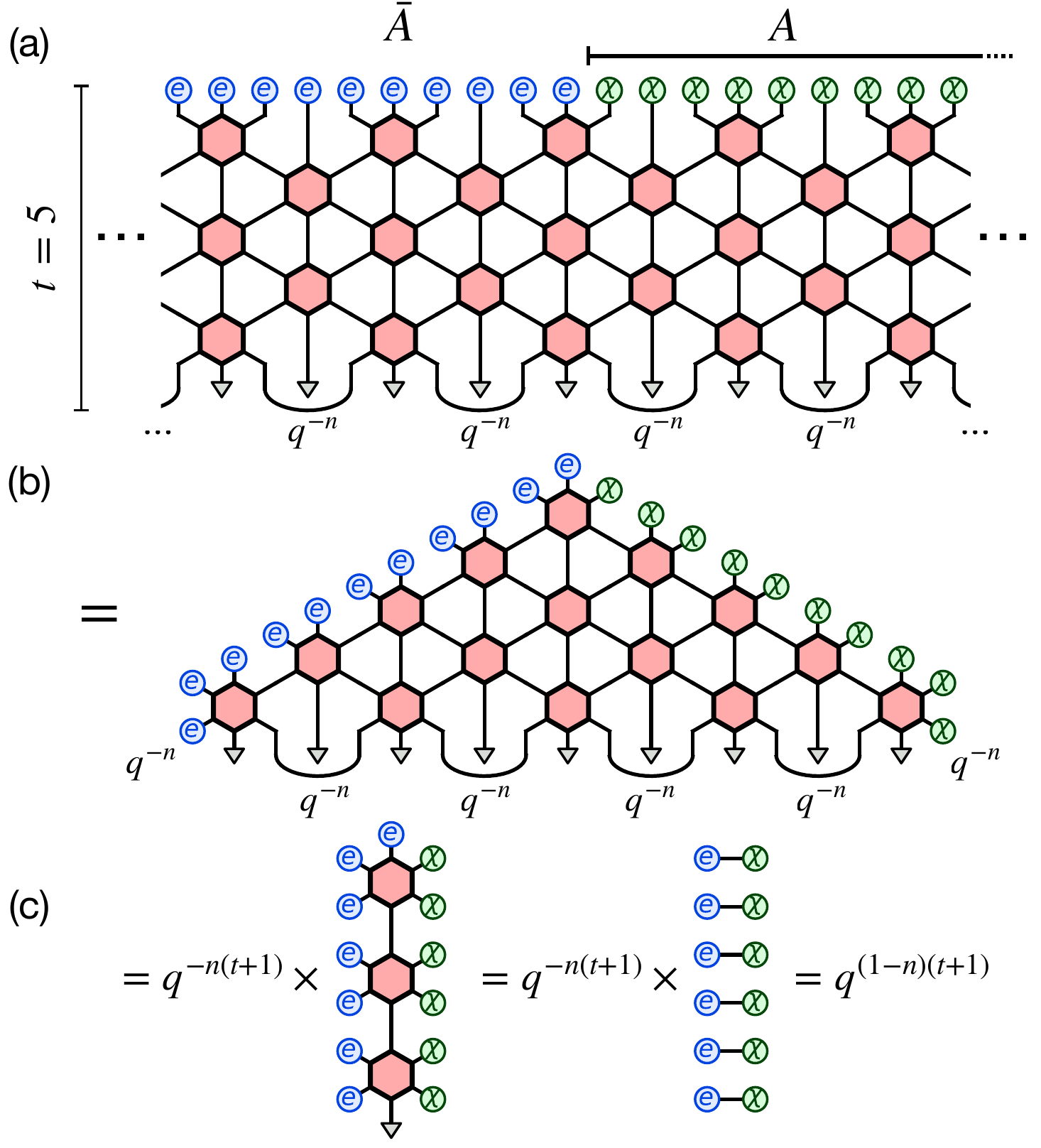}
    \caption{Entanglement growth in tri-unitary circuits: diagrammatic derivation for the $n^\text{th}$ Renyi entropy of a state evolved for time $t=5$.
    (a) Tensor network diagram for $\Tr(\rho_A(t)^n)$. Hexagons represent $n$ replicas of the ``folded'' gate $U\otimes U^\ast$; the initial state (bottom) is an in Eq.~\eqref{eq:solvable_initstate}, with triangles denoting single-qudit $\ket{0}$ states and arcs denoting the Bell pairs $|\bellstate\rangle$ up to the normalization $q^{-n}$ (one factor of $1/q$ for each replica of $|\bellstate\rangle\langle\bellstate|$) indicated explicitly; circles at the top represent contraction of the $n$ ``ket'' and $n$ ``bra'' legs according to the identity permutation $e$ (in $\bar{A}$) or the cyclic permutation $\chi$ (in $A$). 
    (b) Using only unitarity of $U$, all gates outside the backward light cone of the entanglement cut are elided.
    Tri-unitarity and the choice of a ``solvable'' initial state allow further progress: 
    gates can be elided from the bottom right (left) corner by using unitarity of $\tilde{U}$ ($\breve{U}$).
    (c) Gates at the entanglement cut are elided last, resulting in $t+1$ direct contractions between pertmutations $(e|\chi) = q$. The net result is $\Tr\rho_A(t)^n = q^{(1-n)(t+1)}$, thus the $n^{\rm th}$ Renyi entropy (in units of $\ln(q)$) is $S_n = t+1$ for all integers $n\geq 2$.
    }
    \label{fig:entanglement_diagram}
\end{figure}

The derivation is illustrated diagrammatically in Fig.~\ref{fig:entanglement_diagram}, for the case of an entanglement cut $x_\text{cut}$ between two qubits that have both been acted on by the last layer of unitary gates (half of possible cuts are of this kind; the other half are reduced to this kind by eliding the last layer, which does not affect entanglement, and setting $t\mapsto t-1$ in the following derivation). 
$\rho_A^n$ consists of $n$ replicas of the doubled circuit $\mathcal{U} \otimes \mathcal{U}^\ast$: one ``ket'' and one ``bra'' per replica. 
The tensor legs at the final time step $t$ must be contracted appropriately: each ``ket'' leg is paired to a ``bra'' leg according to a permutation $\pi$ in the symmetric group on $n$ elements\cite{Nahum_2018, Vasseur2019}.
In $\bar{A}$ (which is traced out to produce $\rho_A$), each ``ket'' leg is contracted to the ``bra'' leg from the same replica, i.e. the pairing is given by the trivial (or identity) permutation, denoted by $e$. 
In $A$ instead, each ``ket'' leg is contracted with the ``bra'' leg in the \emph{following} replica (in order to implement the product $\rho_A^n$), giving a cyclic permutation $\chi$.

As a consequence of tri-unitarity, whenever a ``stack'' of gates $(U\otimes U^\ast)^{\otimes n}$ has three identical permutations $g^{\otimes 3}$ ($g=e$ or $\chi$) on three adjacent legs, the gates can be elided, and the permutations $g^{\otimes 3}$ moved over to the three output legs.
By using unitarity of $U$ alone, one can elide the circuit everywhere outside the backward light cone of the entanglement cut, turning the tensor network of Fig.~\ref{fig:entanglement_diagram}(a) into that of Fig.~\ref{fig:entanglement_diagram}(b).
We note that contraction between a permutation $e$ or $\chi$ and one of the single-qudit initial states on the odd qudits is simply $\Tr \ket{0}\bra{0}^{\otimes n} = 1$; similarly all Bell pairs contained entirely outside the backward light cone give 1. 
Crucially, Bell pairs that straddle the light cone carry a permutation ($e$ or $\chi$) into an input leg for the gates at the bottom corners of Fig.~\ref{fig:entanglement_diagram}(b): this is the reason for the choice of ``solvable'' initial state in Eq.~\eqref{eq:solvable_initstate}.
Then, unitarity of $\tilde{U}$ and $\breve{U}$ allows further gate elisions starting from the corners and iterating all the way to the cut, leaving only a one-dimensional column of gates, Fig.~\ref{fig:entanglement_diagram}(c). 
Keeping track of the initial state normalization, we find $\Tr \rho_A^n = [(e|\chi) q^{-n}]^{t+1}$, where $(e|\chi)$ denotes the contraction of the two permutations\footnote{In general one has $(\sigma|\tau) = q^{|\sigma^{-1}\tau|}$, where $|g|$ denotes the number of cycles in the permutation $g$. In our case, $g = \chi$ contains a single cycle.}: 
$$
(e|\chi) = \sum_{i_1,\dots i_n=1}^q \langle i_1,\dots i_n | i_2,\dots i_n,i_1\rangle = \sum_{i=1}^q 1 = q \;.
$$
We conclude $\Tr \rho_A(t)^n = q^{(1-n)(t+1)}$, i.e. $S_n = t+1$ (in units of $\log q$).
For the other kind of entanglement cut (adjacent to a qubit that has \emph{not} been acted on by a gate in the last layer), the same derivation yields $S_n(t) = t$.
(Notice that at $t=0$ this cut dependence correctly reduces to whether or not one of the Bell pairs in the initial state straddles the cut.) 
Thus at a fixed cut in space, entanglement alternately grows by 2 (if a gate acts across the cut) and 0 (otherwise), for an average entanglement velocity of exactly 1.

For a finite subsystem with two edges, the derivation proceeds unchanged for each entanglement cut, giving $S_n (t) = 2t+c$ ($c=0$, $1$ or $2$ depending on the cut locations and the parity of $t$ as explained above), as long as $t<\ell/4$ ($\ell$ being the number of qubits of the subsystem). 
After this point, the backward lightcones emanating from each cut intersect, and it is not possible to elide gates further based on tri-unitarity alone. 
Thus $S_n (t)$ is guaranteed to grow at the maximal speed only up to $t\simeq \ell/4$, when it satisfies $S_n(t) \simeq \ell/2$; after that time the behavior may change.
In fact it is easy to find an extreme example in which entanglement growth abruptly stops at $t= \ell/4$: if the odd sublattice is entirely decoupled from the even one, the tri-unitary circuit breaks up into a dual-unitary circuit and a set of inert, disentangled qubits; then, at time $t = \ell/4$, the even sublattice saturates to maximal entanglement ($\ell/2$), while the odd sublattice remains disentangled.

The above results hold for all integer $n \geq 2$. This however implies that they also hold for non-integer $n$, as well as for $n\to 1$ (von Neumann entropy). One can see this as follows.
Considering a finite subsystem $A$ for simplicity, let $\rho_A$ be the reduced density matrix and $\{\lambda_\alpha:\ \alpha=1,\dots q^{|A|} \}$ be its eigenvalues. Our results state that
\begin{equation}
\Tr\rho_A^n = \sum_\alpha \lambda_\alpha^n = q^{(1-n) S}\;,
\end{equation}
for integer $n\geq 2$, where ${S}$ is an $n$-independent integer value, as derived earlier.
The above can be rewritten as 
\begin{equation}
\sum_\alpha (q^S \lambda_\alpha)^n = q^S \;.
\end{equation}
For the sum to stay finite as $n\to\infty$, we see that $q^S \lambda_\alpha \leq 1$ must hold for all $\alpha$; moreover, to match the right-hand side when $n\to\infty$, exactly $q^S$ of the entanglement eigenvalues (note that $S$ is an integer) must satisfy $q^S\lambda_\alpha = 1$, while all others must vanish. 
This constraints the reduced density matrix to the form $\rho_A = P/q^S$, where $P$ is a projector of rank $q^S$: thus the entanglement spectrum is flat and all the entropies (Renyi or von Neumann) coincide.
Based on this fact we simply refer to `entropy' in the following.

In generic tri-unitary circuits, one expects entanglement to saturate to a thermal (infinite-temperature) volume-law given by the Page value\cite{Page_1993} 
$S_\text{Page} = \ell -2^{2\ell-L-1}/\ln(2)$ (in bits) for a subsystem of $\ell$ qubits in a one-dimensional chain of length $L\geq 2\ell$, up to corrections exponentially small in $\ell$. 
Thus entanglement should continue to increase even after the maximum-velocity growth regime stops; however, the behavior becomes model-dependent.

\begin{figure}
    \centering
    \includegraphics[width=1\columnwidth]{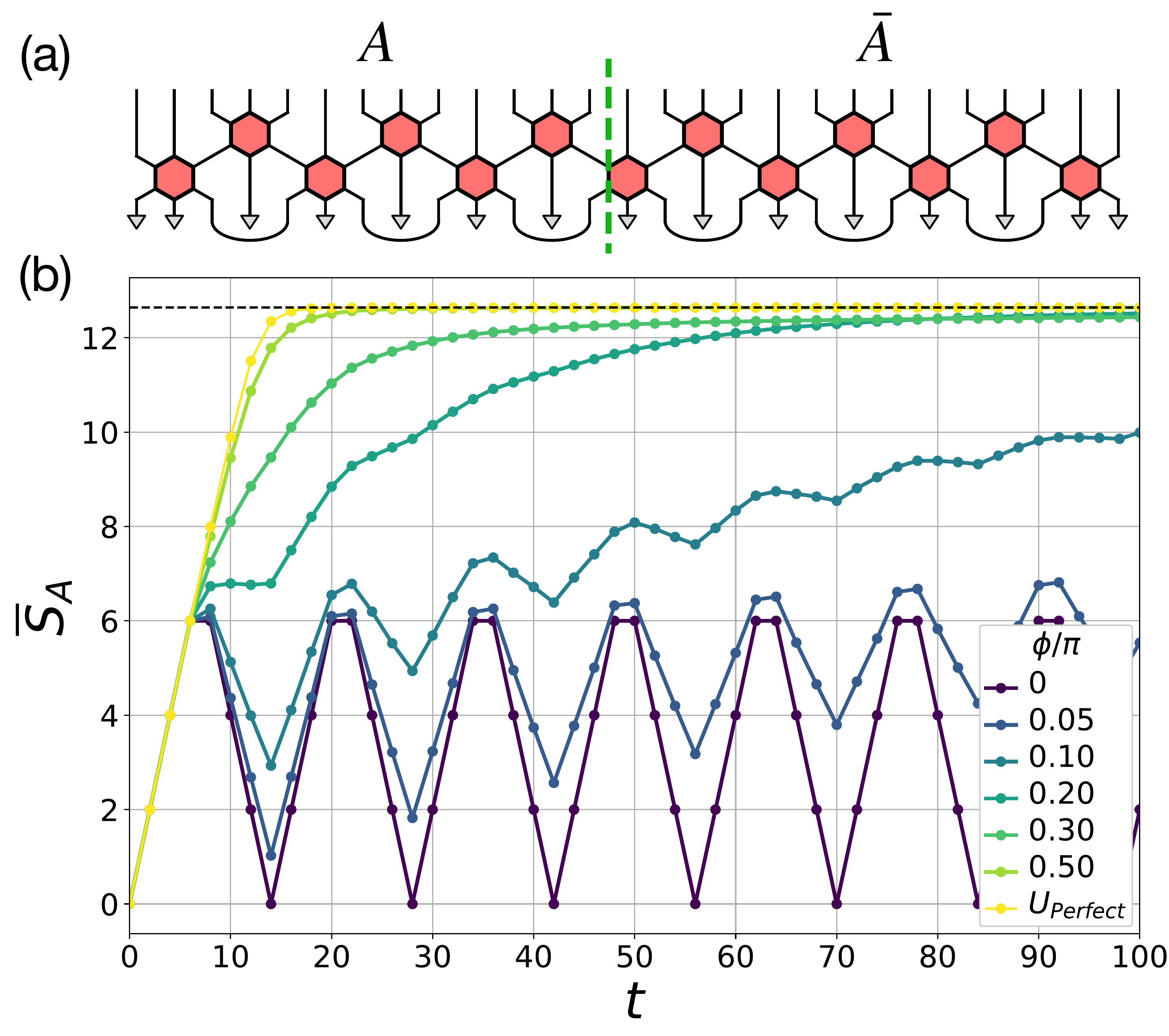}
    \caption{Numerical results for entanglement growth in Floquet tri-unitary circuits.
    (a) Circuit layout: a subsystem $A$ of 13 qubits in a chain of length $L=27$, with open boundary conditions and initial state as in Eq.~\eqref{eq:solvable_initstate}.
    (b) Numerical results for the second Renyi entropy of subsystem $A$, $S_A$ (measured in bits), from exact time evolution.
    The gates are $U_\text{t.u.}[\boldsymbol{\phi}]$ from Eq.~\eqref{eq:UCP} with $\phi_i \equiv \phi$, variable $\phi$, and Haar-random single-qubit gates (data averaged over 10 realizations).
    For $\phi=0$, the model is non-interacting and the 6 Bell pairs in the initial state move ballistically, bouncing off the walls and causing $S_A$ to periodically oscillate between 0 and 6 bits. 
    As interactions are turned on ($\phi >0$), the entropy growth up to $t=6$ remains exactly unchanged, as expected; 
    past $t=6$, the oscillations gradually give way to ballistic growth. 
    We also show data for a circuit made of perfect tensors (dressed with random single qubit gates).
    The dashed line denotes the Page value for this bipartition, $S_\text{Page} = |A|-\frac{1}{4\ln 2}$. }
    \label{fig:ee_growth}
\end{figure}

This phenomenology is illustrated in Fig.~\ref{fig:ee_growth}, which shows the results of exact numerical simulations of entanglement growth under Floquet tri-unitary circuits, for a subsystem $A$ of $\ell = 13$ qubits in a chain of length $L=27$, starting from the solvable initial state Eq.~\eqref{eq:solvable_initstate}.
The circuit setup is sketched in Fig.~\ref{fig:ee_growth}(a);
we choose open boundary conditions, so that the maximum entanglement velocity is 1 rather than 2.
We find that all the circuits, as predicted, exhibit unit entanglement velocity up to $S=6$ bits (the integer part of $\ell/2$); after that point, the curves for different circuits visibly split up, Fig.~\ref{fig:ee_growth}(b).
At $\phi=0$ we have a non-interacting $\SWAP$ circuit, where all entanglement is due to the ballistic motion of the six Bell pairs present in the initial state; 
with an infinite bath $\bar{A}$, the entropy of $A$ would plateau at 6 bits forever, however finite $L$ induces periodic oscillations between 0 and 6 bits (depending on how many Bell pairs straddle the entanglement cut at any given time). 
Adding weak interactions $\phi>0$, the oscillating behavior dictated by the motion of Bell pairs gradually morphs into a steady ballistic trend.
Finally, at strong interactions $\phi\sim\pi/2$ we observe fast growth all the way up to the Page value, though the entanglement velocity becomes sub-maximal immediately after $t=6$.
The same behavior is seen in a Floquet circuit made of perfect tensors (with random single-qubit gates ensuring the circuit is non-Clifford).

Finally, we note that while the non-interacting behaviour ($\phi=0$ in Fig.\ref{fig:ee_growth}) may be captured in dual-unitary circuits with suitable initial states, the interacting behaviour ($\phi>0$ in Fig.\ref{fig:ee_growth}) is unique to tri-unitary circuits. This is because in dual-unitary circuits the entropy must grow linearly until saturation, whereas in tri-unitary circuits the entropy must only grow linearly until half the saturation value. 
Along with the results on correlation functions in Sec.~\ref{sec:correlations}, this is an example of how tri-unitariy circuits give rise to a richer variety of behaviors than dual-unitary ones, while at the same time retaining some of the analytical tractability.

%%%%%%%%%%%%%%%
% HIGHER D
%%%%%%%%%%%%%%%

\section{\label{sec:higher_d}Higher dimension}

Having used tri-unitarity to derive results on correlations and entanglement in $(1+1)$ dimensions, it is interesting to ask about possible applications to higher dimensions. 
The most straightforward extension of dual-unitary circuits consists of ``gluing'' $(1+1)$-dimensional dual-unitary circuits in an additional spatial dimension\cite{Suzuki2021}; this however results in highly anisotropic $(2+1)$-dimensional models, with a ``spacetime duality'' transformation that only acts on two of the three dimensions.

A different approach, which reproduces the phenomenology of $(1+1)$-dimensional dual-unitary circuits more faithfully, is to use ``multi-unitary gates'' arranged on the vertices of a hypercubic lattice in any dimension~\cite{Bertini_2019correlations}.
The tri-unitary gates introduced here can be directly employed in one such construction, as we discuss in the following.
Specifically, we present one construction of $(2+1)$-dimensional quantum circuits built out of 3-qubit tri-unitary gates (as defined in Sec.~\ref{sec:tu_gates}), and exactly derive their two-point correlation functions.
Like in the $(1+1)$-dimensional case, correlations are confined to three special rays; however, in this case the three rays are not co-planar. 
Having three valid arrows of time thus offers the possibility of genuine, intrinsically $(2+1)$-dimensional tri-unitary circuits.

\begin{figure}
    \centering
    \includegraphics[width=\columnwidth]{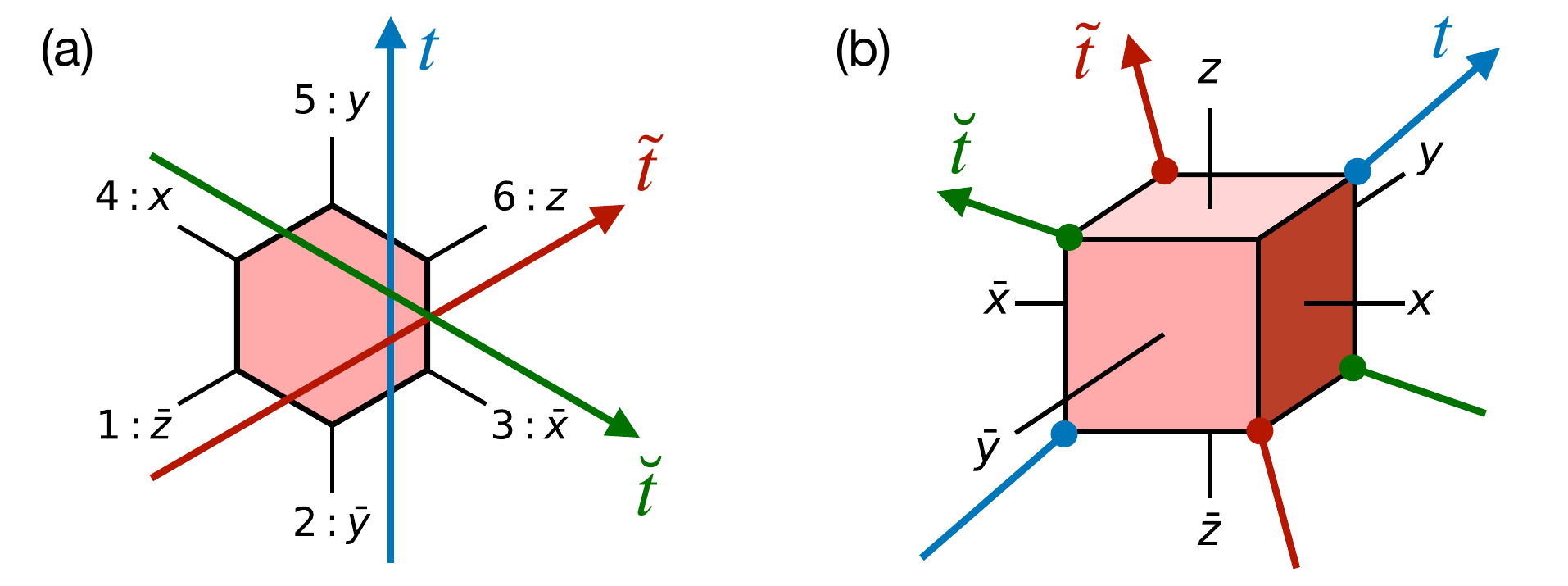}
    \caption{(a) Tri-unitary tensor represented as a hexagon in $(1+1)$-dimensional spacetime; each leg is associated to a vertex. 
    Each contiguous bipartition of legs (associated to one of the three arrows of time $t$, $\tilde{t}$, $\breve{t}$) defines a unitary gate.
    (b) The same tensor viewed as a cube in $(2+1)$-dimensional spacetime; each leg is associated to a face and labelled by the direction perpendicular to it. 
    Each arrow of time crosses the cube through a pair of opposite vertices.
    The correspondence between legs in the two panels is indicated by the labels in (a).
    }
    \label{fig:cube}
\end{figure}

We consider a cubic lattice, $\mathbb{Z}^3$, with a tri-unitary gate $U$ at each vertex. 
Each leg of $U$ connects to one of its 6 nearest neighbors. 
It is convenient to name the legs of $U$ as $x,y,z$ and $\bar{x}, \bar{y}, \bar{z}$, based on which neighbor they connect to, see Fig.~\ref{fig:cube}; 
tri-unitarity of $U$ means that the maps 
$(\bar{x},\bar{y},\bar{z})\mapsto (x,y,z)$, 
$(x,\bar{y},\bar{z})\mapsto (\bar{x},y,z)$, and
$(x,y,\bar{z})\mapsto (\bar{x},\bar{y},z)$ are unitary.
These define three equally valid arrows of time: ${\bf t} \propto (1,1,1)$, $\tilde{\bf t} \propto (-1,1,1)$, and $\breve{\bf t} \propto (-1,-1,1)$. Each of these crosses the cube through a pair of diagonally opposite vertices\footnote{Unitarity of the mapping $(\bar{x},y,\bar{z}) \mapsto (x,\bar{y},z)$, corresponding to the last pair of vertices of the cube, is not assumed.}, as shown in Fig.~\ref{fig:cube}(b).

How do these constraints affect infinite-temperature correlation functions?
Generalizing the idea in Fig.~\ref{fig:overlapping_lightcones} to the present context, we see that each arrow of time restricts two-point correlations to two octants, or tetrahedral light cones: e.g., for two operators separated by a vector $(x,y,z)$, unitarity along ${\bf t}$ requires $x,y,z\geq 0$ (future light cone) or $x,y,z\leq 0$ (past light cone) for connected correlators not to vanish. The other six octants correspond to ``spacelike separations'' relative to ${\bf t}$ and thus cannot have correlations.
Similarly, unitarity along $\tilde{\bf t}$ restricts correlations to two distinct octants, $-x,y,z\leq 0$ or $-x,y,z\geq 0$; and analogously for $\breve{\bf t}$.
Thus unitarity about all three arrows of time limits correlations to \emph{lines}:
namely the ${\bf x}$, ${\bf y}$, and ${\bf z}$ rays (which are the intersection of all three tetrahedral light cones).

\begin{figure}
    \centering
    \includegraphics[width=\columnwidth]{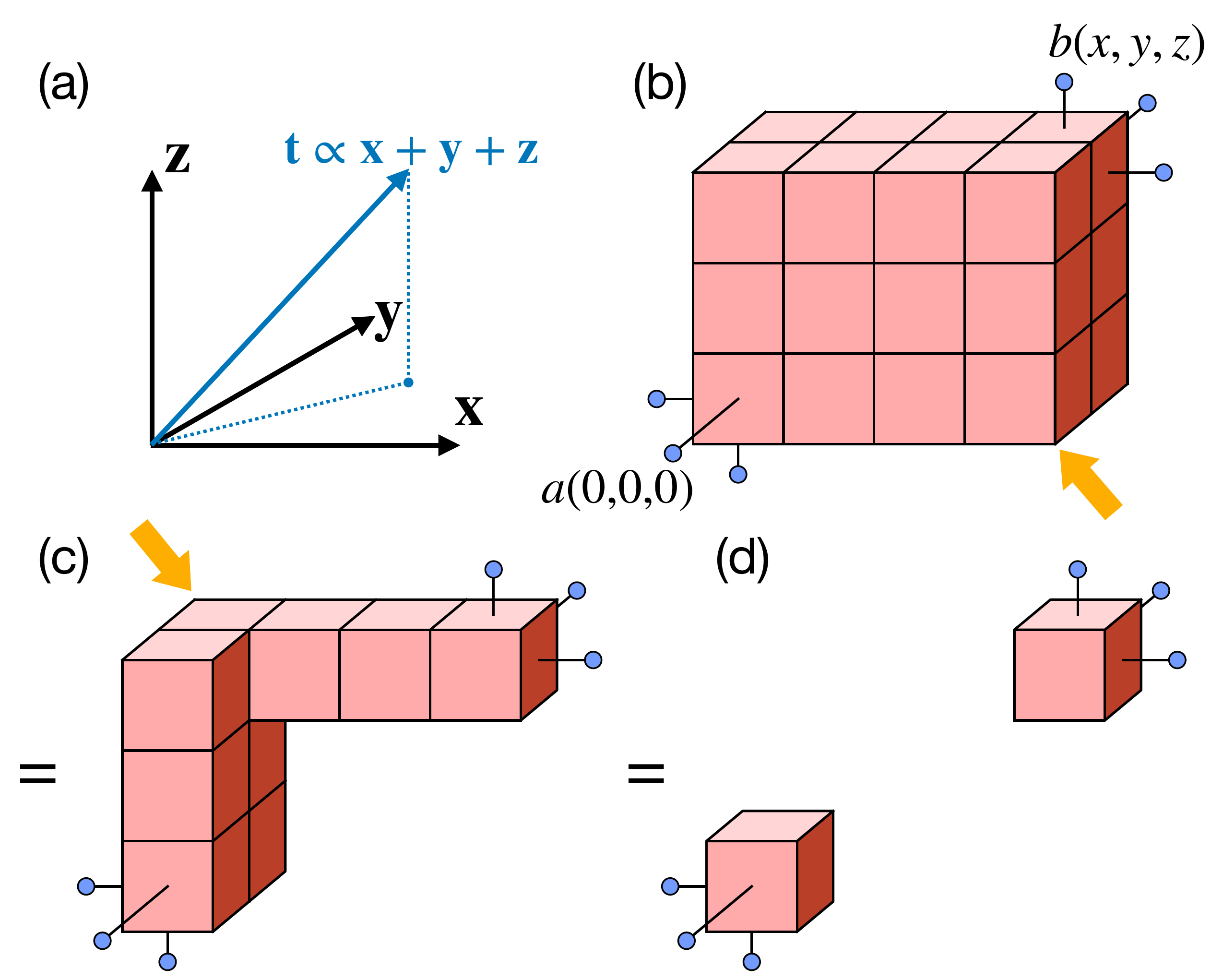}
    \caption{Two-point correlation functions vanish {inside} the light cone for $(2+1)$-dimensional tri-unitary circuits.
    (a) The physical arrow of time is ${\bf t} \propto (1,1,1)$. 
    (b) Tensor network contraction for the infinite-temperature correlation $\langle a(0,0,0) b(x,y,z)\rangle$. Cubes represent ``folded'' tri-unitary gates $U \otimes U^\ast$; legs contracted to non-trivial operators ($a$ and $b$) are shown, while all other exposed faces are implicitly contracted to the identity. 
    Unitarity of $U$ (along $\bf t$) simplifies the tensor network to a hyperrectangle -- the intersection of $a$'s forward light cone and $b$'s backward light cone.
    Unitarity of $\tilde{U}$ causes the elision of gates that have an exposed $(1,-1,-1)$ vertex (starting from the bottom right arrow). 
    (c) This process can be iterated until no more vertices of that kind remain exposed. Then, unitarity of $\tilde{U}^\dagger$ elides all gates with vertex $(-1,1,1)$ exposed (starting from the top left arrow).
    (d) Iteration of this step results in disconnected tensors, and thus a vanishing connected correlator.
    }
    \label{fig:3d_corr}
\end{figure}

\begin{figure}
    \centering
    \includegraphics[width=\columnwidth]{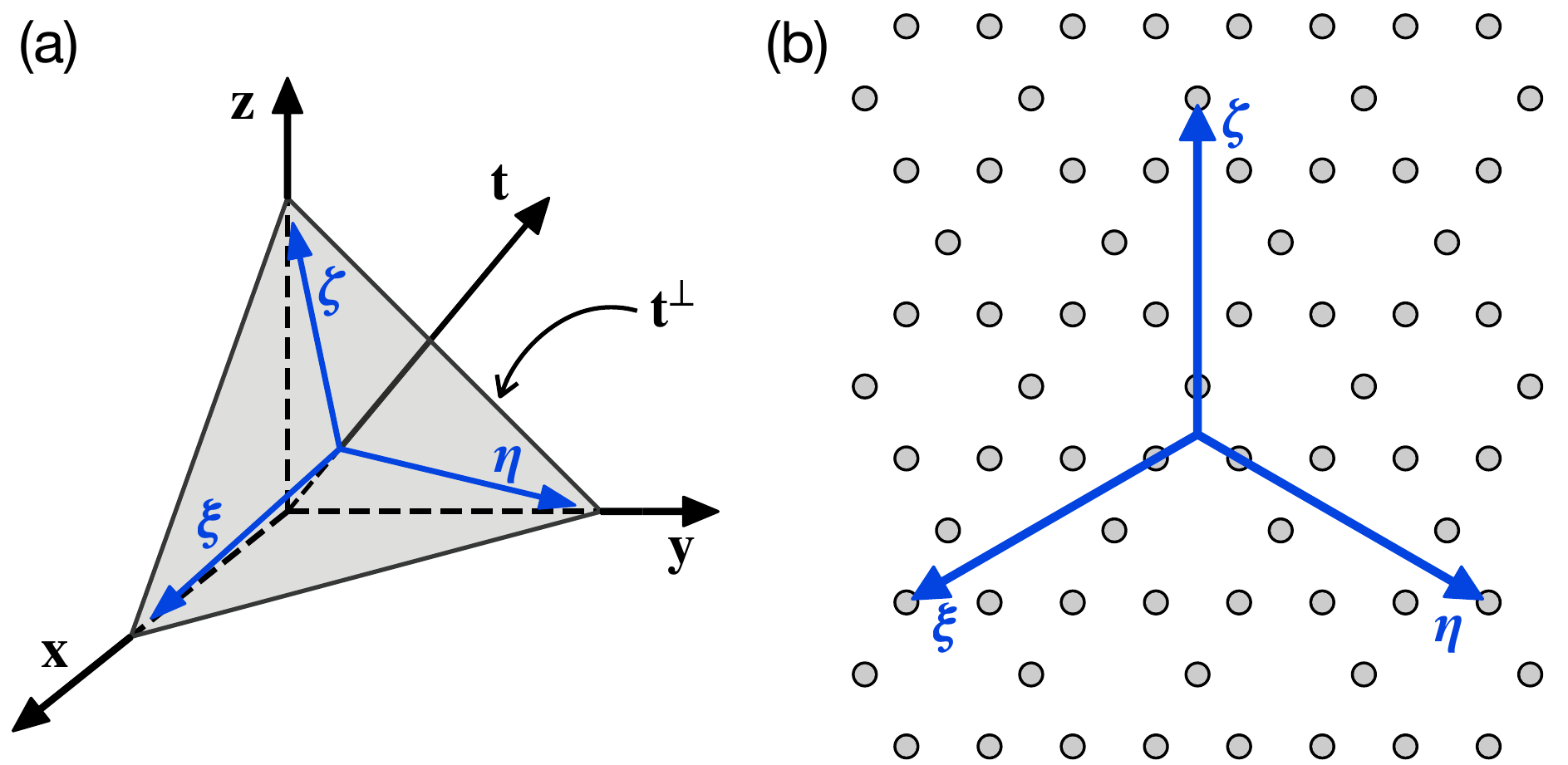}
    \caption{Correlations in $(2+1)$-dimensional tri-unitary circuits. (a) Correlations vanish everywhere except for the rays ${\bf x}$, ${\bf y}$ and ${\bf z}$.
    The arrow of time ${\bf t} \propto {\bf x} + {\bf y} + {\bf z}$ defines spatial planes ${\bf t}^\perp$; the directions $\boldsymbol{\xi}$, ${\boldsymbol \eta}$ and ${\boldsymbol \zeta}$ on the plane are defined by projecting ${\bf x}$, ${\bf y}$ and ${\bf z}$.
    (b) Snapshot of a spatial plane. Qubits live on the sites of a kagome lattice (circles), and two-point correlations propagate ballistically at maximum velocity along the high-symmetry directions $\boldsymbol{\xi}$, ${\boldsymbol \eta}$ and ${\boldsymbol \zeta}$ on the lattice.
    }
    \label{fig:3dgeometry}
\end{figure}

More rigorously, this result can be derived by analyzing the tensor network contraction that corresponds to the correlation function. 
In Fig.~\ref{fig:3d_corr} we derive in detail the fact that correlators vanish at all points \emph{strictly inside} the light cone, $x,y,z>0$.
This only requires unitarity of $U$ and of one between $\tilde{U}$ and $\breve{U}$, which is a less restrictive condition than tri-unitarity.
This less-restrictive condition allows correlations on some surfaces at the boundary of the light cone, e.g. the quadrant $x=0$, $y,z>0$ (where the derivation of Fig.~\ref{fig:3d_corr} would not carry through without invoking unitarity of $\breve{U}$).
On the other hand, full tri-unitarity rules out all correlations except for the three lines ${\bf x}$, ${\bf y}$, ${\bf z}$;
there, the same tensor network analysis shows that correlations are given by iteration of the same quantum channels $M_{\pm, 0}$ found in the $(1+1)$-dimensional case, Eq.~\eqref{eq:M-}, \eqref{eq:M0}, \eqref{eq:M+}. 
The channels do not describe left/right/non-``movers'', as they did in the $(1+1)$-dimensional case of Sec.~\ref{sec:correlations}, but rather three equivalent directions in space: thre projections of ${\bf x}$, ${\bf y}$, ${\bf z}$ onto the spatial plane ${\bf t}^\perp \equiv \{x+y+z = 0\}$, as sketched in Fig.~\ref{fig:3dgeometry}(a).
These three directions in the plane, wich we call $\boldsymbol{\xi}$, $\boldsymbol{\eta}$ and $\boldsymbol{\zeta}$,  form $2\pi/3$ relative angles and are high-symmetry lines in the underlying lattice structure -- a kagome lattice (see Appendix~\ref{app:2d}), shown in Fig.~\ref{fig:3dgeometry}(b).
Correlations travel along $\boldsymbol{\xi}$, $\boldsymbol{\eta}$ and $\boldsymbol{\zeta}$ at the same (maximal) velocity, and vanish everywhere else.

This phenomenology generalizes that of $(1+1)$-dimensional dual- and tri-unitary circuits in an interesting way. 
Correlations are not pinned to manifolds of co-dimension 1 (the surface of a suitable light cone), as one may have guessed, but rather to manifolds of dimension 1 (lines).
This is particularly striking in non-ergodic models, where this result implies that some operators (eigenmodes with $\lambda_{\mu, i} = 1$) move ballistically along one of three special directions in two-dimensional space, Fig.~\ref{fig:3dgeometry}; These directions are picked by the underlying lattice structure and circuit architecture.
This ``subdimensional'' propagation of information is reminiscent of fractonic excitations pinned to subdimensional manifolds\cite{Nandkishore2019, Vijay2015, Pretko2017}, albeit in a dynamical, driven setting, in which there is no meaningful notion of energy and the only quantity moving along subdimensional manifolds is information.

Finally we remark on implementing this construction as a local quantum circuit on two-dimensional qubit arrays. 
This is slightly subtle: the space-like slices of the $(2+1)$-dimensional tensor network described above, i.e. the planes ${\bf t}^\perp$ orthogonal to ${\bf t}$, intersect the qubit worldlines at locations that vary with time over the course of one period (meant here as a period of the circuit architecture; the gates need not be time-periodic). 
Thus it looks like the qubit array itself should change its geometric structure during the dynamics.
However, in Appendix~\ref{app:2d} we show that it is possible to sidestep this issue and implement the desired circuit with local gates on a \emph{static} 2D array of qubits by introducing two ancillas for every system qubit.
The qubits are arranged on a kagome lattice, with system qubits occupying one of three sublattices and ancillas occupying the other two (Fig.~\ref{fig:3dgeometry}(b) shows only the system qubits).
All ancillas are initialized in a fiduciary state, say $\ket{0}$, and are returned to this initial state after each period; meanwhile, the system qubits undergo the desired time evolution\footnote{Quantum circuits enhanced with ancillas in this way are equivalent to the class of quantum cellular automata\cite{Schumacher2004, Arrighi2019, Farrelly2020, Piroli2020qca}, i.e. unitary transformations that preserve the locality of operators but are not necessarily realizable via finite-depth local circuits.}.

%%%%%%%%%%%%%%%
% CONCLUSION
%%%%%%%%%%%%%%%

\section{Conclusion and outlook \label{sec:conclusion}}

The dynamics of many-body quantum systems out of equilibrium is a notoriously hard problem for both theory and computation.
Models that afford a degree of analytical control or solvability are extremely useful in elucidating phenomena and principles that may apply to more general, less tractable scenarios; yet such solvable models are rare.

In this work, we have introduced a new, large family of quantum many-body evolutions, dubbed tri-unitary circuits, in which crucial properties including correlations and entanglement are analytically tractable.
This tractability stems from their peculiar causal structure, which features three distinct ``arrows of time'' under which the dynamics are unitary. This builds on previous results on dual-unitary circuits, whose unitarity under two distinct arrows of time has enabled the derivation of a plethora of exact results.

Tri-unitary circuits generalize and extend the construction of dual-unitary circuits in several important ways. 
The different circuit architecture, featuring three-qubit gates arranged at the vertices of a triangular lattice in spacetime, results in a different symmetry -- rather than exchanging space and time, it mixes the two nontrivially. This has sharp consequences in the phenomenology of these systems: correlations are allowed to propagate along \emph{three} special directions in spacetime, namely the light rays $\delta x = \pm v\delta t$ as well as the static worldline $\delta x=0$. Information can thus move at the ``speed of light'' or not move at all -- the latter a qualitatively different possibility absent in dual-unitary circuits.
While tri-unitary circuits are expected to be strongly chaotic in general, it is intriguing to speculate that this feature (the presence of strictly non-moving operators) might inspire constructions of tractable circuit models of localization~\cite{Chandran_2015}.
A richer phenomenology also arises in the growth of quantum entanglement. In tri-unitary circuits (starting from a class of ``solvable'' initial states), entropy grows ballistically at the maximal velocity, but only up to an entropy density of half the maximum, at which point the behavior may change based on the model (in contrast with dual-unitary circuits where the maximum-velocity growth persists up to maximum density).

Another novel aspect of tri-unitarity is the possibility of genuine higher-dimensional extensions. 
Higher-dimensional constructions of dual-unitary circuits\cite{Suzuki2021} are highly anisotropic: having two co-planar arrows of time, they are effectively stacks of coupled $(1+1)$-dimensional dual-unitary layers\footnote{This definition of dual-unitary circuits, based on the number of unitary arrows of time, excludes the higher-dimensional generalizations mentioned in Ref.~[\onlinecite{Bertini_2019correlations}]}.
In other words, the action of spacetime duality by necessity exchanges time and \emph{one} spatial direction, leaving out any others\cite{Grover2021}. 
On the contrary, we have constructed tri-unitary circuits in $(2+1)$-dimensional spacetime that treat all dimensions on the same footing, due to the presence of three non-coplanar arrows of time.
Interestingly, correlations in these circuits are pinned to three special light-rays -- correlations propagate at maximal velocity along three high-symmetry directions of the underlying lattice, at $2\pi/3$ angles with each other. The propagation of information along sub-dimensional manifolds is in itself a novel feature of these circuits, reminiscent of quasiparticles with sub-dimensional mobility in fractonic systems\cite{Vijay2015, Pretko2017, Nandkishore2019}, though in a completely different (driven, non-equilibrium) context.

Our work opens several directions for future research. 
First, while we have provided a large (31-parameter) family of tri-unitary gates on qubits, it would be interesting to obtain a full parametrizations of all tri-unitary gates (whether on qubits or higher-dimensional qudits).
Second, regarding the phenomenology of these dynamics, it would be interesting to obtain more general results on the approach to thermalization, the growth of entanglement from generic (non-``solvable'') initial states, and other diagnostics of quantum chaos such as the spectral form factor and out-of-time-ordered correlators.
Regarding circuit architectures, we have argued that tri-unitary circuits saturate the number of possible ``arrows of time'' in flat $(1+1)$-dimensional spacetime, due to the absence of regular lattices with higher symmetry; however, more exotic generalizations may be possible, e.g. via quasicrystalline tilings or on curved spaces -- the latter potentially connecting to ideas in quantum gravity\cite{Pastawski_2015, Hayden2016}, as well as recent explorations of quantum simulation in curved spaces\cite{Kollar2019, Boettcher2020}.
Finally, we note that the triangular circuit structure is not invariant under ``spacetime duality'' -- a $\pi/2$ rotation maps the circuit to a sequence of non-local matrix-product operators (with finite bond dimension represented by a spacelike qubit worldline). 
This poses a challenge in deriving results about the spectral form factor of these circuits, as it makes the transfer matrix employed in Ref.~[\onlinecite{Bertini_2018sff}] non-unitary and non-local.
On the other hand, this may present opportunities for the study of non-unitary dynamics via spacetime duality\cite{Ippoliti_2021postselection, Ippoliti_2021fractal, Grover2021}, by allowing access to potentially distinctive types of non-unitary evolutions involving matrix-product operators rather than local circuits.

\begin{acknowledgments}
We acknowledge useful discussions with Tibor Rakovszky, Yuri Lensky, and Bruno Bertini. This work was supported with funding from the Defense Advanced Research Projects Agency (DARPA) via the DRINQS program (M.I.), the Sloan Foundation through a Sloan Research Fellowship (V.K.) and by the US Department of Energy, Office of Science, Basic Energy Sciences, under Early Career Award No. DE-SC0021111 (V.K. and C.J.). The views, opinions and/or findings expressed are those of the authors and should not be interpreted as representing the official views or policies of the Department of Defense or the U.S. Government. M.I. was funded in part by the Gordon and Betty Moore Foundation's EPiQS Initiative through Grant GBMF8686. Numerical simulations were performed on Stanford Research Computing Center's Sherlock cluster.
\end{acknowledgments}

\bibliography{main}

\onecolumngrid

\appendix

%%%%%%%%%%%%%%%
% EXAMPLES
%%%%%%%%%%%%%%%

\section{Examples of correlations \label{app:CPcorrelations}}

In this appendix we show that a two-parameter family of gates within the parametrization of Eq.~\eqref{eq:UCP} can realize all the correlation behaviors of the ``ergodic hierarchy'' reviewed in Sec.~\ref{sec:du}. We define the gate 
\begin{align}
    U(\phi,g) &=  \SWAP_{3,1}\; \CP_{3,1} (\phi)\; \CP_{2,3}(\phi)\; \CP_{1,2}(\phi) e^{i \phi/2 \sum_i Z_i} e^{-ig \sum_i X_i} \nonumber \\
    & = \SWAP_{3,1} \; e^{-i\frac{\phi}{2} (Z_1 Z_2 + Z_2 Z_3 + Z_3 Z_1)} e^{-ig\sum_i X_i}
\end{align}
which is $U_\text{t.u.}$ from Eq.~\eqref{eq:UCP} with $\phi_i \equiv \phi$, $u_i \equiv v_i \equiv \id$, and $w_i \equiv e^{-igX_i}$.
We can compute the transfer matrices
\begin{align}
    M_\mu = \begin{pmatrix} 1 & 0 & 0 &0 \\
    0 &\cos(\frac{\phi}{2})^2 & 0 & 0 \\
    0 &0 & \cos(2g) \cos(\frac{\phi}{2})^2 & \sin(2g)\\
    0 &0 & -\sin(2g) \cos(\frac{\phi}{2})^2 & \cos(2g) \end{pmatrix}
\end{align}
which are independent of $\mu = \pm,0$ due to the symmetry of the gate.
Varying parameters $\phi$ and $g$, we can realize the entire ``ergodic hierarchy'' of correlation functions:
\begin{itemize}
    \item[(i)] {\bf non-interacting}: setting $g=\phi=0$ yields $U_{(i)}=U(0,0)= \SWAP_{3,1}$. The transfer matrices are $M_\mu = \mathcal{I}$ (the identity channel), thus the eigenvalues
    \begin{align}
    \lambda_\mu^{(i)}&=\{1,1,1,1\}
    \end{align}
    
    \item[(ii)] {\bf interacting, non-ergodic}: setting $\phi \neq 0$ with $g=0$ gives (up to a global phase) $U_{(ii)}=\SWAP_{1,3} e^{-i(\phi/8)Z_\text{tot}^2}$ where $Z_\text{tot} = Z_1+Z_2+Z_3$; thus two-point functions of $Z_i$ remain constant along the rays while those of $X_i$, $Y_i$ decay:
    \begin{align}
        \lambda_\mu^{(ii)}&=\{1,\cos(\phi/2)^2, \cos(\phi/2)^2,1\}
    \end{align}
    
    \item[(iii)] {\bf ergodic, non-mixing}: setting $g = \pi/2$ adds a $\pi$-pulse about the $x$ axis to the previous drive, $U_{(iii)}= \SWAP_{1,3} e^{-i(\phi/8)Z_\text{tot}^2} X_1 X_2 X_3$, causing $Z$ correlations to oscillate:
    \begin{align}
        \lambda_{0,\pm}^{(iii)}&=\{1,-\cos(\phi/2)^2, \cos(\phi/2)^2,-1\}
    \end{align}
    Thus time-averaged $Z$ correlators decay, but instantaneous ones do not.
    
    \item[(iv)] {\bf ergodic, mixing}: obtained for generic values of $g$, $\phi$. We find the eigenvalues
    \begin{align}
        \lambda_\mu^{(iv)}&= \{1,\cos(\phi/2)^2, \frac{1}{4} \cos(2g) (3+\cos(\phi))-\frac{1}{8}f(\phi,g),\frac{1}{4} \cos(2g) (3+\cos(\phi))+\frac{1}{8}f(\phi,g) \}\\
        f(\phi,g) &=  \sqrt{-13-20\cos(\phi)+2\cos(4g)(3+\cos(\phi))^2 + \cos(2\phi))}
    \end{align}
    with eigenoperators $\{\id, X, h(\phi,g)Y \pm Z\}$, with 
    \begin{align}
        h(\phi,g) &= e^{2i(3g+\phi)} f(\phi,g) + 4e^{2i(3g+\phi)} \cos(2g)\sin(\phi/2)^2 \;.
\end{align}
    Thus all correlations decay exponentially, without time averaging.
\end{itemize}

Finally one may add an extreme case, dubbed ``Bernoulli circuits''\cite{Aravinda2021},
These are represented by the perfect tensor\cite{Pastawski_2015}, in which all transfer matrices are erasure channels, $M_\mu(O) = \Tr(O)\id/2$, and thus all correlators decay to 0 immediately. 
This property follows from quantum error correction, namely from the fact that no information about the encoded qubit should be accessible from any single one of the physical qubits. (The $M_\mu$ channels correspond to encoding a logical qubit into 5 physical qubits, discarding (tracing out) 4 of them, and retaining the last one as output).

\begin{figure}
\centering 
\includegraphics{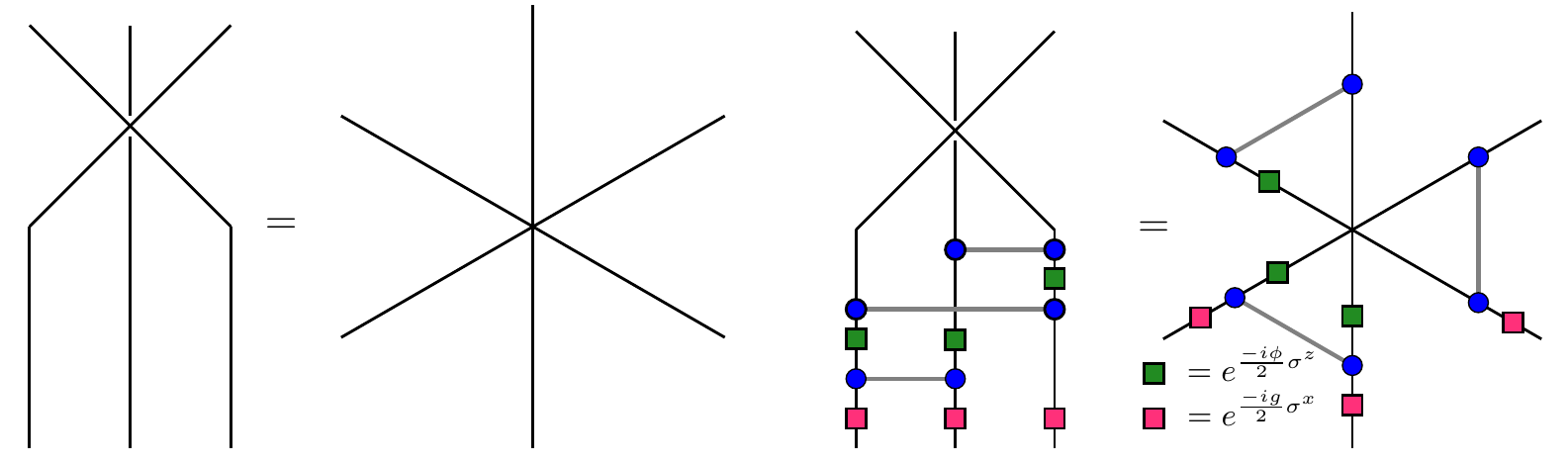}
\caption{Left: case (i), right: case (ii)-(iv)} 
\end{figure}

%%%%%%%%%%%%%%%
% HIGHER D DETAILS
%%%%%%%%%%%%%%%

\section{Details on two-dimensional circuit \label{app:2d}}

Here we discuss the implementation of the 3-dimensional tri-unitary tensor network of Sec.~\ref{sec:higher_d} as a quantum circuit in two spatial dimensions. 

To review, our construction involves a cubic lattice $(x,y,z)\in \mathbb Z^3$ with tri-unitary gates at each vertex; 
the qubit worldlines travel in three directions, ${\bf x}$, ${\bf y}$ and ${\bf z}$ (given by $x\in\mathbb{R}$, $y,z\in\mathbb{Z}$ and permutations thereof), intersecting at gates; 
each gate thus has 3 input qubits and 3 output qubits. 
There are three equally valid axes of time, we take $t = x+y+z$ as the physical or ``laboratory'' time for concreteness.
Expressing the above tensor network as a two-dimensional quantum circuit is not entirely straightforward because the lattice configuration of qubits in the space-like plane ${\bf t}^\perp$ (spanned e.g. by ${\boldsymbol \xi} = (2,-1,-1)/\sqrt{6}$ and $\boldsymbol{\eta} = (-1,-1,2)/\sqrt{6}$) changes during the dynamics: na\"ively, the qubits would have to be physically moved on the plane during the evolution.
However, this can be avoided by introducing ancilla qubits, as we explain in the following.

\begin{figure*}
\centering
\includegraphics[width=\textwidth]{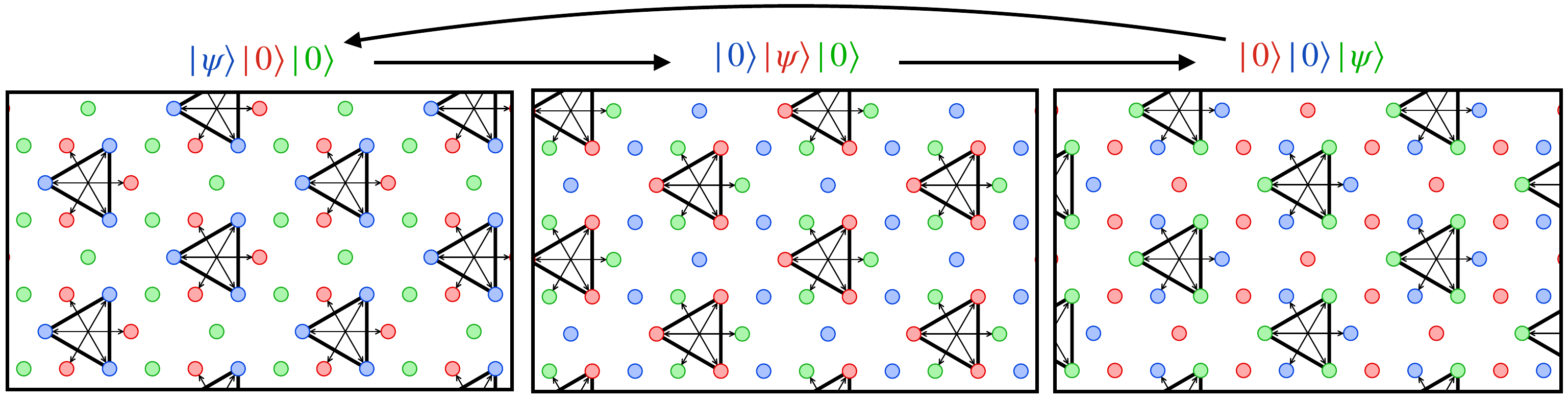}
\caption{$(2+1)$-dimensional tri-unitary circuit implementation in two spatial dimensions. 
Qubits are arranged on a kagome lattice and partitioned into three sublattices (red, green and blue). The state of interest occupies only one sublattice at any given time; the other two store ancillas in a fixed product state, e.g. $\ket{0}^{\otimes 2N}$. 
A floquet cycle consists of three layers of unitary gates. 
Before the first layer (left), the state $\ket{\psi}$ lives in the blue sublattice. Tri-unitary gates (thick lines) couple triplets of blue qubits as shown; then, $\SWAP$ gates (thin arrows) act on pairs of blue and red qubits, moving the state to the red sublattice. 
The second (center) and third (right) layers of unitary gates are applied similarly. The state of interest changes sublattice at each layer, and returns to the blue sublattice after three layers (a Floquet period, if the gates are time-independent). \label{fig:qca}}
\end{figure*}

Let us for simplicity assume all gates $U$ in the circuit are equal; the circuit then is time-periodic, with a Floquet period of $t=3$: $(x,y,z)\mapsto (x+1,y+1,z+1)$ is the shortest translation along the temporal direction $\bf{t}$ that leaves the lattice invariant. 
Three layers of unitary gates take place within each Floquet period -- at times $t=3n$, $3n+1$, and $3n+2$. 
In between two layers of unitary gates, say at $t = 3n+k+1/2$, the location of qubits on the spatial plane is given by intersecting the qubit worldlines with the planes $x+y+z = k+1/2$, which yields a kagome lattice for any $k \in \mathbb{Z}_3$.
However, the three kagome lattices are distinct: namely the blue sublattice in Fig.~\ref{fig:qca} for $k = 0$, the red one for $k=1$, and the green one for $k=2$.
The union of these three lattices is itself a kagome lattice.

In order to implement the $(2+1)$-dimensional tri-unitary dynamics as a local circuit in 2-dimensional space, we place a physical qubit on every site of the kagome lattice obtained above; 
however, the state of interest $\ket{\psi}$ is stored only on the blue sublattice, while the rest of the qubits (red and green sublattices) are ancillas initialized in a trivial product state, say $\ket{0}^{\otimes 2N}$.
We then perform the tri-unitary gates of Eq.~\eqref{eq:UCP} by acting on triplets of blue sites with controlled-phase gates (thick lines in Fig.~\ref{fig:qca}), plus any single-qubit rotations;
then we swap each qubit with the diametrically opposite vertex of the hexagonal plaquette where the gate has acted (arrows in Fig.~\ref{fig:qca}).
As a result, the blue sites are now occupied by trivial states $\ket{0}^{\otimes N}$, while the state of interest is written on the red sites. 
This process implements the $t=0$ layer of unitary gates. 
The $t=1$ and 2 layers are implemented analogously, as shown in the other panels of Fig.~\ref{fig:qca}, with the state of interest moving to the green sites and finally back to the blue sites, completing a Floquet cycle.
Because after a period the state of the ancillas is unchanged, this evolution belongs to the $\textsf{QC}_a$ class of quantum circuits augmented by ancillas.
This class is equivalent to quantum cellular automata\cite{Schumacher2004, Arrighi2019, Farrelly2020, Piroli2020qca}. 
We note that in this case there is no fundamental obstruction to realizing the evolution as a low-depth local circuit, merely a technical inconvenience (the three-qubit gates would have to act on triplets of qubits separated by several lattice spacings, as opposed to the present case where all interactions take place around a plaquette).

If all interactions are turned off, the system evolves by a sequence of $\SWAP$ gates along three special directions; 
it is clear then that opereators propagate ballistically, at fixed velocity, along one of three high-symmetry directions in the kagome lattice, at $2\pi/3$ angles with each other.
These are the projections of ${\bf x}$, ${\bf y}$ and ${\bf z}$ on the ${\bf t}^\perp$ plane.
More surprisingly, this phenomenology is robust to the addition of arbitrary tri-unitary interactions, except correlators on the special rays generically decay exponentially in time rather than being 1 -- their behavior is again dictated by the quantum channels $M_{0,\pm}$, as in the $(1+1)$-dimensional case.

\end{document}